\pdfoutput=1
\documentclass[usenatbib]{mnras}
\usepackage[T1]{fontenc}
\usepackage{ae,aecompl}
\usepackage{url}	
\usepackage{amssymb}	
\usepackage{graphicx}
\usepackage{subfigure}
\usepackage{amsmath}
\usepackage{color}
\usepackage{multirow}
\usepackage{multicol}
\usepackage{threeparttable}
\usepackage{verbatim}
\usepackage{tikz}
\usepackage{float}
\usepackage{todonotes}
\usepackage{algorithm,algpseudocode}
\usepackage{aas_macros}
\usepackage{ae,aecompl}
\hypersetup{colorlinks=true,linkcolor=red,citecolor=blue,breaklinks=true}
\usepackage[english]{babel}
\usepackage[normalem]{ulem}
\usepackage{array}
\usepackage[capitalise,noabbrev]{cleveref}

\usepackage{adjustbox}
\usepackage{gensymb}
\usepackage{makecell}

\def\deg{\ifmmode^\circ\else$^\circ$\fi}
\def\pdeg{\ifmmode $\setbox0=\hbox{$^{\circ}$}\rlap{\hskip.11\wd0 .}$^{\circ}
          \else \setbox0=\hbox{$^{\circ}$}\rlap{\hskip.11\wd0 .}$^{\circ}$\fi}

\def\mo{\ifmmode^{-1}\else$^{-1}$\fi}
\def\,{\thinspace}
\def\arcm{$^{\scriptstyle\prime}$}

\def\,{\thinspace}

\def\GHz{\ifmmode $\,GHz$\else \,GHz\fi}
\def\mKs{\ifmmode $\,mK\,s$^{1/2}\else \,mK\,s$^{1/2}$\fi}
\def\muKs{\ifmmode \,\mu$K\,s$^{1/2}\else \,$\mu$K\,s$^{1/2}$\fi}
\def\MJysr{\ifmmode \,$MJy\,sr\mo$\else \,MJy\,sr\mo\fi}

\def\KJysr{\ifmmode \,$kJy\,sr\mo$\else \,kJy\,sr\mo\fi}
\def\MJysrmK{\ifmmode \,$MJy\,sr\mo$\,mK$_{\rm CMB}\mo\else \,MJy\,sr\mo\,mK$_{\rm CMB}\mo$\fi}
\def\microns{\ifmmode \,\mu$m$\else \,$\mu$m\fi}
\def\muK{\ifmmode \,\mu$K$\else \,$\mu$\hbox{K}\fi}
\def\microK{\ifmmode \,\mu$K$\else \,$\mu$\hbox{K}\fi}
\def\muW{\ifmmode \,\mu$W$\else \,$\mu$\hbox{W}\fi}
\def\kms{\ifmmode $\,km\,s$^{-1}\else \,km\,s$^{-1}$\fi}
\def\cm2{\ifmmode $\,cm$^{-2}\else \,cm$^{-2}$\fi}
\def\mum{\ifmmode $\,\mu$m\else \,$\mu$m\fi}

\DeclareMathAlphabet{\mathsc}{OT1}{cmr}{m}{sc}
\def\testbx{bx}%
\DeclareRobustCommand{\ion}[2]{%
\relax\ifmmode
\ifx\testbx\f@series
{\mathbf{#1\,\mathsc{#2}}}\else
{\mathrm{#1\,\mathsc{#2}}}\fi
\else\textup{#1\,{\mdseries\textsc{#2}}}%
\fi}

\newcommand{\healpix}{\ensuremath{\tt HEALPix}}

\newcommand{\smica}{\ensuremath{\tt SMICA}}

\newcommand{\kcmb}{\ensuremath{{\rm K}_{\rm CMB}}}

\newcommand{\NHI}{\ensuremath{N_\mathsc {Hi}}}

\newcommand{\hi}{\ensuremath{\mathsc {Hi}}}
\newcommand{\hid}{\ensuremath{\mathsc {\hi\text{-d}}}}
\newcommand{\Nside}{\ensuremath{N_{\rm side}}}

\newcommand{\calH}{\mathcal{H}}
\newcommand{\calN}{\mathcal{N}}
\newcommand{\calP}{\mathcal{P}}
\newcommand{\pd}{\partial}

\graphicspath{{./fig/}}

\title[Bayesian inference of dust emissivity]{Bayesian inference methodology to characterize the dust emissivity at far-infrared and submillimeter frequencies}

\author[Adak et al.]{Debabrata Adak $^{1,2,3,4}$\thanks{E-mail: adak@iac.es},
Shabbir Shaikh$^{5}$\thanks{sshaik14@asu.edu},
Srijita Sinha$^{6}$,
Tuhin Ghosh$^{6}$,
\newauthor Francois Boulanger$^{7}$, Guilaine Lagache$^{8}$, Tarun Souradeep$^{9,10,4}$ and
\newauthor Marc-Antoine Miville-Desch\^enes$^{11}$ 
\\
$^{1}$ Instituto de Astrofísica de Canarias, E-38200 La Laguna, Tenerife, Spain\\
$^{2}$ Departamento de Astrofísica, Universidad de La Laguna, E-38206 La Laguna, Tenerife, Spain\\
$^{3}$ The Institute of Mathematical Sciences, CIT Campus, Tharamani, Chennai, Tamil Nadu 600113, India\\
$^{4}$Inter University Centre for Astronomy and Astrophysics, Post Bag 4, Ganeshkhind, Pune-411007, India\\
$^{5}$ School of Earth and Space Exploration, Arizona State University, Tempe, AZ 85287, USA\\
$^{6}$ National Institute of Science Education and Research, An OCC of Homi Bhabha National Institute, Bhubaneswar 752050, Odisha, India\\
$^{7}$ Laboratoire de Physique de l’Ecole normale sup\'erieure, ENS, Universit\'e PSL, CNRS, Sorbonne Universit\'e, Université Paris Cit\'e, F-75005 Paris, France\\
$^{8}$ Aix Marseille Univ., CNRS, CNES, LAM, Marseille, France\\
$^{9}$ Raman Research Institute, C. V. Raman Avenue, Bengaluru, Karnataka 560080, India\\
$^{10}$ Indian Institute of Science Education and Research, Dr. Homi Bhabha Road, Ward No. 8, NCL Colony, Pashan, Pune-411008, India \\ 
$^{11}$ AIM, CEA, CNRS, Universit\'e Paris-Saclay, Universit\'e de Paris, F-91191 Gif-sur-Yvette, France
}

\date{Accepted XXX. Received YYY; in original form ZZZ}

\pubyear{2024}

\begin{document}
\label{firstpage}
\pagerange{\pageref{firstpage}--\pageref{lastpage}}
\maketitle

\begin{abstract}
We present a Bayesian inference method to characterise the dust emission properties using the well-known dust-\hi\ correlation in the diffuse interstellar medium at Planck frequencies $\nu \ge 217$\,GHz.  We use the Galactic \hi\ map from the Galactic All-Sky Survey (GASS) as a template to trace the Galactic dust emission. We jointly infer the pixel-dependent dust emissivity and the zero level present in the Planck intensity maps. We use the Hamiltonian Monte Carlo technique to sample the high dimensional parameter space ($D \sim 10^3$). We demonstrate that the methodology leads to unbiased recovery of dust emissivity per pixel and the zero level when applied to realistic Planck sky simulations over a 6300 deg$^2$ area around the Southern Galactic pole. As an application on data, we analyse the Planck intensity map at 353 GHz to jointly infer the pixel-dependent dust emissivity at $\Nside=32$ resolution (1.8\deg\ pixel size) and the global offset. We find that the spatially varying dust emissivity has a mean of 0.031 $\MJysr (10^{20} \mathrm{cm^{-2}})^{-1}$ and $1\sigma$ standard deviation of 0.007 $\MJysr (10^{20} \mathrm{cm^{-2}})^{-1}$. The mean dust emissivity increases monotonically with increasing mean \hi\ column density. We find that the inferred global offset is consistent with the expected level of Cosmic Infrared Background (CIB) monopole added to the Planck data at 353 GHz. This method is useful in studying the line-of-sight variations of dust spectral energy distribution in the multi-phase interstellar medium. 
\end{abstract}



\section{Introduction} \label{sec:1}
Galactic dust emission is one of the significant foregrounds in measurements of the Cosmic Microwave Background (CMB) intensity and polarization at frequencies above $\sim$100 GHz \citep{planck-IV:2018}. At this frequency range, the form of spectral energy distribution (SED) of the Galactic dust and the Cosmic Infrared Background (CIB,  \citealt{Puget:1996, Lagache:2005sw}) are similar. While interesting in its own right, characterisation and understanding of the Galactic dust properties are crucial for measuring the CMB B-mode polarization \citep{planck-XI:2018}. To mitigate this foreground contribution from the measurements, it is essential to understand how dust grains contribute to the B-mode polarisation. Understanding the Galactic dust emission is also critical for the reliable reconstruction of the CIB anisotropies \citep{planck-XLVIII:2016}. The fact that the CIB and the Galactic dust share similar spectral properties makes it crucial to understand the spatial distribution and the SED of Galactic dust to separate the two emissions reliably. The CIB traces the matter distribution in the Universe and plays a crucial role in delensing the lensed B-mode contribution in the observed B-mode measurements \citep{Larsen:2016}. It is also an important probe to be cross-correlated with other probes of the large-scale structure \citep{Planck:2013qqi, ACT:2014swt, Planck:2015emq, Maniyar:2018hfp, 2023MNRAS.520..583J}. Moreover, even though the CIB is weakly polarized, at the frequencies where CIB is significant, the polarization of the CIB can be a contaminant to CMB B-mode polarization \citep{Lagache:2019xto, Feng:2019miy}. 

Various methods have been used to study the nature of the Galactic dust emission with the Modified Black Body (MBB) as a model of the Galactic dust SED at Planck\footnote{\url{http://www.esa.int/Planck}} frequencies. Inferring Galactic dust SED parameters at the highest Planck angular resolution with sufficient signal-to-noise ratio is also important because smoothing of resolution may reduce information associated with the Galactic dust. \cite{planck-X:2015} use \texttt{Commander} method to perform Bayesian inference of the Galactic dust spectral properties at 7.5\arcm\ full-width half maximum (FWHM) angular resolution. A complementary method to do a similar task is implementing the dust-\hi\ correlation. Galactic dust is correlated with \hi\ 21 cm line emission of neutral hydrogen at high Galactic latitude and low-column density regions \citep{Boulanger:1988}. Hence, the Galactic \hi\ emission map can be used as a tracer of the Galactic dust emission. Using Planck and IRAS data along with \hi\ 21 cm observations obtained by Green Bank Telescope (GBT), \citealt{planck-XXIV:2011} estimate the dust emissivities in 14 fields covering more than 800 $\rm{deg}^2$ at high Galactic latitude. Using the same formalism, \cite{planck-XVII:2014} study the dust emissivity and its SED by fitting the MBB model over the Southern Galactic Pole (SGP, $b < -25\deg$). A proper estimation of the Galactic dust emission is needed to separate the CIB emission (see, e.g., \citealt{planck-XLVIII:2016, Lenz:2019, Irfan:2019}). Once the contribution of Galactic dust to a given frequency map is estimated, it can be subtracted to obtain the contribution of the CIB anisotropy \citep{planck-XXX:2014}. \cite{Lenz:2019} have produced the CIB maps by cross-correlating \hi4PI data \citep{hi4pi2016aa} with the Planck intensities at 353, 545, and 857 \GHz\ over approximately 25$\%$ of the sky.

Planck does not measure the absolute sky background referred to as the zero levels of intensity, which have been fixed at the level of map-making \citep{planck-VIII:2013, planck-VIII:2015}. Two considerations go into determining the zero level of an HFI frequency map: the zero level of the Galactic emission and the CIB monopole. Zero level of the Galactic emission has been set using dust-\hi\ correlation at high Galactic latitude where \hi\ is the reliable Galactic dust tracer. The underlying assumption is that the dust emission is zero where \hi\ column density is zero. The offset at 857 \GHz\ is obtained by cross-correlating the Planck HFI data at 857 \GHz\ with Leiden Argentine Bonn (LAB) Galactic \hi\ Survey data. For other HFI frequencies, the Galactic zero level has been fixed using cross-correlation of maps at respective frequencies with the 857 \GHz\ map (for details, see section 5.1 of \citealt{planck-VIII:2013}). The estimated offsets are subtracted from the detector data at the time of map-making. After this step, the CIB monopole, estimated from the CIB model of \cite{B_thermin:2012}, has been added to each Planck HFI frequency intensity map.

Separation of the Galactic dust emission from the Planck frequency maps needs to take into account the proper treatment of the offset present in the maps. In the previous studies, the dust emissivities are fitted over local sky patches and variable offsets. \cite{planck-XXIV:2011} utilises dust-\hi\ correlation to estimate emissivity properties of the dust at Low, Intermediate, and High-Velocity Clouds (LVC, IVC, and HVC) over 14-fields. The analysis in \cite{planck-XVII:2014} estimates the emissivity at LVC and offset using dust-\hi\ correlation over the patches of 15\deg\ diameter. Both works use the $\chi^2$ minimisation technique. When dealing with very high dimensional parameter problems, $\chi^2$ minimisation often results in parameter estimates that are far from the typical set \citep{Mackay:2003}. 
Furthermore, the efficiency of the $\chi^2$ minimisation method is limited by the correlation between the parameters of interest and becomes inefficient for the high dimensional parameter space. \cite{planck-XI:2013} measures global offset first and then estimates the Galactic dust SED parameters by fitting the MBB spectrum to offset-corrected intensity maps. In this work, we utilise the dust-\hi\ correlation to jointly infer the dust emissivity and the \textit{global offset} of the whole sky region considered instead of individual smaller sky patches. We use GASS \hi\ data with Planck intensity maps over the same sky region used in the study of \cite{planck-XVII:2014}. We show that such an analysis can benefit from the joint inference of the emissivity and the global offset.

We sample the joint posterior distribution of emissivity and the global offset. The total number of variables to sample in this problem is around $2 \times 10^{3}$. We use the Hamiltonian Monte Carlo (HMC) sampling method, which can more efficiently sample such large dimensional parameter space than the Metropolis-Hastings (MH) algorithm \citep{Duane:1987de}. For both HMC and MH algorithms, a new proposed point ($\theta^{\star})$ is generated from the present point ($\theta)$ by taking some sort of steps based on the target density. For a $D$-dimensional problem, the number of steps required to reach a nearly independent proposal point grows as $D^{1/4}$ for HMC and $D$ for MH algorithm. The total amount of computation time with a reasonable acceptance rate grows as $D^{5/4}$ for HMC and $D^2$ for MH algorithm \citep{Creutz:1988,Neal:2012,Betancourt:2013}. HMC uses the gradient of the posterior distribution to generate a proposed point, which, in principle, has an acceptance probability equal to one. This feature has led to HMC being increasingly employed in the high dimensional inference problems encountered in cosmology \citep{Hajian:2007, Taylor_Cl_with_HMC_2008, Jasche_HMC_LSS, Jasche_Wandelt2013, Anderes_Wandelt_Lavaux_bayes_lensing_2015}, including the inference of CMB foreground parameters \citep{Grumitt:2019ape}. 

This paper is structured along the following lines. In \cref{sect:2}, we present the data model, the likelihood, and the details of the HMC sampler. \cref{sec:2} describes the data used in this paper and the pre-processing of the data before the main analysis. Validation of the Bayesian inference method using simulated maps is discussed in \cref{sec:4}. In \cref{sec:5}, we present and discuss the CMB-subtracted Planck 353 GHz intensity map analysis results. We summarise the main results of the paper in \cref{sec:summary}.

\section{Method}\label{sect:2}

This section discusses the data model and likelihood analysis to sample the model parameters from their posterior distributions using the HMC sampler.

\subsection{Dust emission model and the data likelihood}\label{sect:2.1}

We are interested in characterising the dust emission properties in the diffuse interstellar medium over the frequency range covered by the Planck HFI maps ($\nu\,\ge\,217$\,GHz). The major contributors to the total emission in this range of frequencies are CMB, Galactic dust, CIB, and instrumental noise. We assume that the CMB intensity map derived using the well-established component separation techniques (SMICA, NILC, SEVEM, and COMMANDER) from the Planck multi-frequency observations is accurate enough in the sky region not obscured by the Galactic disc \citep{planck-IV:2018}. With this assumption, we work with the CMB-subtracted Planck frequency maps to reduce the total number of components that need to be modelled. The \hi\ column density map can be used as a tracer of the dust emission due to strong dust-gas correlation in the diffuse ISM \citep{Boulanger:1988}. We model the CMB-subtracted intensity map ($d_{\nu}$) at frequency $\nu$ as the addition of the \hi-correlated dust emission ($I^{\hid}$), the noise and a constant global offset. In general, the noise term can consist of the CIB emission $(I^{\rm CIB}_{\nu})$, the instrument noise ($I^{\rm N}_{\nu}$), and the Galactic residual emission ($I^{\rm R}_{\nu}$). The Galactic residuals contain the dust emission from $\texttt{H}_2$ gas, which is uncorrelated with the \hi-emission. The global offset ($O_{\nu}$) over the regions being analysed includes the contribution mainly from the CIB monopole and emission not accounted for by the \hi\ emission. 

It is evident from previous studies that the dust emissivity varies smoothly over the sky \citep{planck-XVII:2014}. This fact allows us to assume the emissivity to be constant over a set of pixels. This set of pixels is determined using the \healpix\ grid at a coarser resolution (i.e. lower \Nside) than the resolution of the data map. The bigger pixel over which emissivity is assumed constant is termed a \emph{superpixel}, whereas pixels within the superpixel are called \emph{subpixels}. This model is expressed in the following equation,
\begin{equation}
\label{eq.1}
    d_{\nu}(\Omega^{j}_{i}) = I^{\hid}_{\nu}(\Omega^j_{i}) + O_{\nu} + I^{\rm N}_{\nu}(\Omega^j_{i}) + I^{\rm CIB}_{\nu}(\Omega^j_{i}) + I^{\rm R}_{\nu}(\Omega^j_{i}) ,
\end{equation}
where $\Omega^j_i$ indicates the direction of $i^{th}$ subpixel within $j^{th}$ superpixel. We model the \hi-correlated dust emission $(I^{\hid}_{\nu})$ as
\begin{equation}
\label{eq:dust_model}
I^{\hid}_{\nu}(\Omega^j_{i}) = \epsilon^j_{\nu} \NHI(\Omega^j_{i}) \ , 
\end{equation}
where \NHI\ denotes the integrated column density of \hi\ component in the direction $\Omega^j_i$, and $\epsilon^j_{\nu}$ is the dust emissivity at frequency $\nu$ at $j^{th}$ superpixel. Here, we consider a single \hi\ template to trace the \hi-correlated dust emission. Because we are treating $I^{\rm CIB}_{\nu}$ and $I^{\rm R}_{\nu}$ as noise along with $I^{\rm N}_{\nu}$, we call the remaining contribution of $I^{\hid}_{\nu}$ and $O_{\nu}$ as \textit{the signal model}, $s_{\nu}$,
\begin{equation}\label{eq:model}
s_{\nu}(\Omega^j_{i}) = \epsilon^j_{\nu} \NHI(\Omega^j_{i}) + O_{\nu} \ .
\end{equation}
We assume that the three noise components, the CIB, the Galactic residuals and the instrument noise, are independent.
Hence, the covariance matrix of the total noise $(\boldsymbol{\Sigma_{\nu}})$  at frequency $\nu$ is the sum of the covariance of the individual noise component. We denote the angle between $i^{\rm th}$ subpixel of $j^{\rm th}$ superpixel and the $i'^{\rm th}$ subpixel of $j'^{\rm th}$ superpixel by $\theta^{jj'}_{ii'}$ and the elements of the covariance matrix are denoted by $\Sigma^{jj'}_{\nu, ii'} \equiv \Sigma_{\nu}(\theta^{jj'}_{ii'})$.  While we consider the correlation between the subpixels within a superpixel, we neglect the correlation between subpixels that belong to different superpixels, that is $\Sigma^{jj'}_{\nu, ii'} = \Sigma^{jj}_{\nu, ii'} \delta_{jj'}$. Hence, the nonzero elements of $\boldsymbol{\Sigma_{\nu}}$ are given by
\begin{equation}\label{eq:cov_mat}
\Sigma^{jj}_{\nu, ii'} = \Sigma_{\nu}^{\rm N}(\theta^{jj}_{ii'}) + \Sigma_{\nu}^{\rm CIB}(\theta^{jj}_{ii'}) + \Sigma_{\nu}^{\rm R}(\theta^{jj}_{ii'}),
\end{equation}
where $\boldsymbol{\Sigma_{\nu}^{\rm N}}$ denotes the instrument noise covariance, $\boldsymbol{\Sigma_{\nu}^{\rm CIB}}$ and $\boldsymbol{\Sigma_{\nu}^{\rm R}}$ denote the contribution to $\boldsymbol{\Sigma_{\nu}}$ due to the CIB and the Galactic residuals, respectively. Unlike instrument noise, $I_{\nu}^{\rm CIB}$ and $I_{\nu}^{\rm R}$ are spatially correlated signals. Hence, its contribution to the total noise covariance matrix gives rise to the non-zero off-diagonal terms in $\boldsymbol{\Sigma_{\nu}}$. We further assume the instrument noise, the CIB and the residual Galactic emission to be Gaussian with their respective covariance. Hence, the joint likelihood of all the data elements given the model parameters $(\epsilon^j_{\nu}, O_{\nu})$ is
\begin{equation}\label{eq_likelihood}
\mathcal{L}(\{d_{\nu}(\Omega^j_i)\} | \{\epsilon^j_{\nu}, O_{\nu} \}) = \frac{1}{(2\pi)^{D/2} \sqrt{|\boldsymbol{\Sigma_{\nu}}|} } \exp{\Big[ -\frac{\chi^2}{2}\Big ] },
\end{equation}
where $\chi^2$ is
\begin{equation}\label{eq_Chi_square}
\chi^2 = \sum_{j} \sum_{i, i' \subset j} [d_{\nu}(\Omega^j_i) - s_{\nu}(\Omega^j_i)] [\Sigma_{\nu}^{-1}]^{jj}_{ii'} [d_{\nu}(\Omega^j_{i'}) - s_{\nu}(\Omega^j_{i'})].
\end{equation}
$[\Sigma^{-1}]^{jj}_{ii'}$ represents the element of the inverse of the covariance matrix $\boldsymbol{\Sigma}$. 

In the model fitting, templates $\NHI(\Omega^j_{i})$ and the noise variance $\Sigma_{\nu}$ are known, and $\{\epsilon^j_{\nu}, O_{\nu} \}$ are the unknown parameters of the model that we aim to infer from the observed data. Bayes theorem allows us to write the \textit{posterior probability distribution} ($\calP( \{\epsilon^j_{\nu}, O_{\nu} \} | \{d_{\nu}(\Omega^j_{i})\})$) of the parameters given data as
\begin{equation}\label{posterior}
    \calP( \{\epsilon^j_{\nu}, O_{\nu} \} | \{d_{\nu}(\Omega^j_{i})\}) = \frac{\mathcal{L}(\{d_{\nu}(\Omega^j_{i})\} | \{\epsilon^j_{\nu}, O_{\nu} \}) \calP(\{\epsilon^j_{\nu}, O_{\nu} \}) }{ \calP(\{d_{\nu}(\Omega^j_{i})\})},
\end{equation}
where $\calP(\{\epsilon^j_{\nu}, O_{\nu} \})$ is the \textit{prior probability distribution} of the parameters, and $\calP(\{d_{\nu}(\Omega^j_{i})\})$ is the \textit{evidence}. We assume a uniform prior for all the parameters of interest without any bounds. Because the model is linear in parameters and we assume a Gaussian likelihood, the posterior distribution is also a Gaussian as a function of the model parameters. $\calP(\{d_{\nu}(\Omega^j_{i})\})$ acts as a normalisation constant. Hence, the functional dependence of the posterior distribution on parameters is the same as that of the likelihood distribution up to a proportionality constant. We sample the posterior distribution given in \cref{posterior} to get the joint samples of all the parameters of interest $\{\epsilon^j_{\nu}, O_{\nu} \}$. We have around $2 \times 10^3$ emissivity parameters per \NHI\ template and one offset parameter.

\subsection{Additional terms in the modelling}\label{subsec:add_term}
In this section, we discuss additional terms that may be required in modelling the data, their motivation, and implementation in the inference methodology.
\subsubsection{Dipole term}\label{subsubsec:dipole_term}
We can model and fit for a dipole contribution in the signal model,
\begin{equation}\label{eq:model_with_dipole}
    s_{\nu}(\Omega^j_{i}) = \epsilon^j_{\nu} \NHI(\Omega^j_{i}) + O_{\nu} + D_{\nu}(\Omega^j_{i}),
\end{equation}
where $D_{\nu}(\Omega^j_{i})$ accounts for residual dipole due to the CMB dipole, the CIB dipole, and the dipole from the Galactic residuals. In harmonic space, the expression for the dipole is
\begin{equation}\label{eq:dipole_expr}
    D_{\nu}(\Omega^j_{i}) = a^{\nu}_{1,0} Y_{1,0}(\Omega^j_{i}) + a^{\nu}_{1,1} Y_{1,1}(\Omega^j_{i}) + a^{\nu}_{1,-1} Y_{1,-1}(\Omega^j_{i}),
\end{equation}
where $Y_{1,m}$ are spherical harmonics with $a^{\nu}_{1,m}$ being spherical harmonic coefficients.
Using the relations $Y_{l,-m} = (-1)^m Y^*_{l,m} $ and $a_{l,-m} = (-1)^m a^*_{l,m}$, the above expression can be rephrased in terms of $m = 0$ and $m = 1$ coefficients as:
\begin{equation}\label{eq:dipole_expr_1}
    D_{\nu}(\Omega^j_{i}) = a^{\nu}_{1,0} Y_{1,0}(\Omega^j_{i}) + 2 a^{R, \nu}_{1,1} Y^R_{1,1}(\Omega^j_{i}) - 2 a^{I, \nu}_{1,1} Y^I_{1,1}(\Omega^j_{i}),
\end{equation}
where superscripts $R$ and $I$ indicate real and imaginary parts of a complex quantity, respectively. In the inference process, it is convenient to treat the dipole in harmonic space because with the Gaussian likelihood, posterior is also Gaussian as a function of $a_{1,m}$ due to their linear nature, unlike the real space variables indicating dipole amplitude and the direction of the dipole.

\subsubsection{Multiple templates}\label{subsubsec:multiple_templates}

The dust emission in far-infrared and sub-millimetre frequency bands can also be modelled as a linear combination of multiple \hi\ templates with different dust emissivity per template. For example, \citet{planck-XXIV:2011} estimate the dust emissivity properties associated with LVC, IVC and HVC clouds over 14 fields. \citet{T_Ghosh:2017} and \citet{Adak:2019} estimate the mean dust emissivity over Southern and Northern Galactic pole regions, respectively, by correlating the CMB-subtracted Planck 353 GHz map with \hi\ column density associated with cold, lukewarm and warm neutral medium (CNM, LNM, and WNM). \citet{Lenz:2019} use the individual spectral channel map of \hi\ brightness temperature from the HI4PI survey~\citep{hi4pi2016aa} to model the dust emission at Planck HFI frequencies $\nu \ge 353$\,GHz. The modelling and inference framework used in this work can be extended to include multiple \NHI\ templates with the signal modelled as
\begin{equation}\label{eq:model_multi_t}
s_{\nu}(\Omega^j_{i}) = \sum^{N_t}_{t = 1} \epsilon^{j,t}_{\nu} \NHI^{t}(\Omega^j_{i}) + O_{\nu} + D_{\nu}(\Omega^j_{i}),
\end{equation}
where the summation is over $N_t$ number of \NHI\ templates ($\NHI^{t}$), indexed by $t$, and $\epsilon^{j,t}_{\nu}$ is corresponding emissivity.

We test the impact of additional terms like dipole term or multiple \hi\ templates on the inferred $\{\epsilon^{j,t}_{\nu}, O_{\nu} \}$ parameters using simulated maps at Planck frequencies in \cref{sec:4}.

\subsection{CIB model power spectra}\label{sec:2.3}
\begin{figure}
\includegraphics[width=\columnwidth]{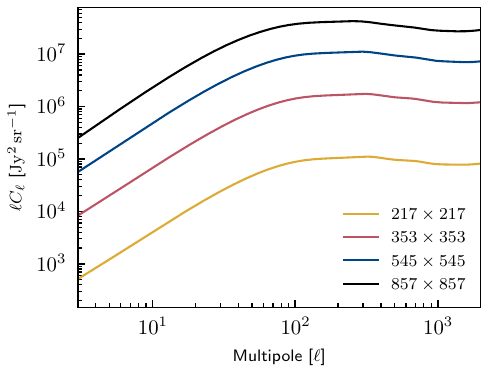}
\caption{The CIB model power spectrum (in units of $\ell C_{\ell}$) obtained by fitting the CIB model including the shot noise at four HFI frequencies \citep{planck-XXX:2014}. We use the model $C_{\ell}^{\rm CIB}$ to compute the full covariance matrix $\Sigma_{\nu}^{\rm CIB}$. }
\label{fig:2}
\end{figure}
The CIB is a relic emission from stellar-heated dust within galaxies formed throughout cosmic history. At far-infrared/sub-millimetre bands and the resolution of Planck, the CIB appears as a diffuse background emission in the total intensity. The CIB anisotropies are found to be correlated across the frequencies and follow approximately $\ell^{-1}$ power law angular power spectrum \citep{planck-XVIII:2011, planck-XXX:2014}. We adopt the best-fit CIB model power spectra (including the shot noise) at 217, 353, 545, and 857 GHz obtained by \cite{planck-XXX:2014}.  \cref{fig:2} presents the model CIB power spectra used in our analysis.

We treat the CIB anisotropies as Gaussian and correlated noise in the Planck intensity maps. To take into account the correlation between pixels at a given smoothing scale, we compute the CIB covariance matrix, $\Sigma_{\nu}^{\rm CIB}$, at a given frequency $\nu$ between two pixels $i$ and $i'$ using the relation
\begin{equation}\label{eq:2.3.1}
 \Sigma_{\nu}^{\rm CIB}(\theta_{ii'}) = \frac{1}{4\pi} \sum (2\ell+1) \, C_{\ell, \nu \times \nu}^{\rm CIB} \, B_{\ell}^2 \, w_{\ell}^2 \, P_{\ell} (\hat{n}_i. \hat{n}_{i'}),
\end{equation}
where $C_{\ell}^{\rm CIB}$ is the CIB power spectrum, $B_{\ell}$ is the beam window function of the Gaussian beam, $P_{\ell}$ is the Legendre polynomial of order $\ell$ and $w_{\ell}$ is the \healpix\ pixel window function.

\subsection{Hamiltonian Monte Carlo sampling} \label{sect:2.2}
In this section, we present the essentials of the Hamiltonian Monte Carlo sampling formalism to draw the samples of $\{\epsilon^j_{\nu}\}$ and $\{O_{\nu} \}$ from the distribution given in \cref{posterior}, which is the same as the likelihood in \cref{eq_likelihood} for a uniform prior on parameters. Details of the HMC sampling method and some considerations that went into analysing the parameter samples are discussed in \cref{app:HMC_details}.

HMC uses Hamiltonian dynamics to generate the proposed point and traverse the parameter space. The method treats the parameters $\{ \epsilon^j_{\nu} \}$ and $\{ O_{\nu} \}$ as position variables and augments them with the momentum variables $(p^{j,t}_{k\nu},\,p_{O_{\nu}})$ to define phase space dynamics.The Hamiltonian of the dynamics for the given problem is
\begin{equation}\label{def_H}
\calH(p^{j,t}_{\nu}, p_{O_{\nu}}, \epsilon^{j,t}_{\nu}, O_{\nu})  = \frac{p^2_{O_{\nu}}}{2\mu_{O_{\nu}}} + \sum_{j} 
\frac{{p^{j,t}}^{2}_{\nu}} {2\mu^{j,t}_{\nu}} - \ln[\calP(\{ \epsilon^{j,t}_{\nu}, O_{\nu}\})],
\end{equation}
where $\calP(\{ \epsilon^{j,t}_{\nu}, O_{\nu}\})$ is the parameter posterior as obtained in \cref{posterior} and $\mu^{j,t}_{\nu}$, $\mu_{O_{\nu}}$ are mass terms for $\epsilon^{j,t}_{\nu}$ and $O_{\nu}$, respectively. Up to a constant, which is independent of parameters of interest, $\ln[\calP(\{ \epsilon^{j,t}_{\nu}, O_{\nu}\}) ]$ is
\begin{equation}
\ln[\calP(\{ \epsilon^{j,t}_{\nu}, O_{\nu}\})]  = - \frac{1}{2} \chi^2 \ ,
\end{equation}
where $\chi^2$ is given by \cref{eq_Chi_square}.

While considering the total covariance matrix, only considering the diagonal part leads to underestimating the parameter uncertainty. Whereas considering all the elements, including the ones corresponding to two subpixels of different superpixels, drastically increases the computation cost. We take the approach somewhere in between. We consider the correlations between subpixels corresponding to a given superpixel only and neglect the correlations among the pixels of two different superpixels. This is expressed in \cref{eq:cov_mat}. The mathematical expressions given in the subsequent discussion are under this approximation.

We need the \textit{time derivatives} of position and momentum variables to simulate the Hamiltonian dynamics. The time derivative of the position corresponding to a given parameter is simply momentum divided by the mass of the corresponding parameter:
\begin{equation}
 \dot{\epsilon}^{j,t}_{\nu} =  \frac{\pd \calH }{\pd p^{j,t}_{\nu}} = 
\frac{p^{j,t}_{\nu}}{\mu^{j,t}_{\nu}} \quad \text{and} \quad
\dot{O}_{\nu} =  \frac{\pd \calH }{\pd p_{O_{\nu}}} = 
\frac{p_{O_{\nu}}}{\mu_{O_{\nu}}},
\end{equation}
which is not at all a computationally involved operation. The time derivative of the momentum involves computing the derivative of the logarithm of the posterior:
\begin{equation}
 \dot{p} \equiv \frac{\pd \calH}{\pd q} = -\frac{\pd \ln(\calP)}{\pd q}.
\end{equation}
Our model is linear in the parameters of interest, and the likelihood is a Gaussian distribution. Hence, with the flat priors, the derivatives of the logarithm of the posterior are simple expressions given below. The derivative with respect to emissivity is
\begin{equation}\label{P_dot_eps_rev}
 \frac{\pd \ln(\calP)}{\pd \epsilon^{j,t}_{\nu}} = 
\sum_{i,i' \subset j} \NHI^{t} (\Omega^j_{i}) [\Sigma_{\nu}^{-1}]^{jj}_{ii'} [ d_{\nu} - 
s_{\nu} ]^j_{i'} \ ,
\end{equation}
and derivative with respect to offset $O_{\nu}$ is 
\begin{equation}\label{P_dot_Onu_rev}
 \frac{\pd \ln(\calP)}{\pd O_{\nu}} = 
\sum_{j} \sum_{i,i' \subset j} [\Sigma_{\nu}^{-1}]^{jj}_{ii'} [d_{\nu} -  
s_{\nu} ]^j_{i'}.
\end{equation}

While sampling the model parameters, including the residual CMB dipole contribution, the spherical harmonic coefficients $a^{(R/I), \nu}_{1m}$ in \cref{eq:dipole_expr_1} are jointly sampled along with emissivity and offset as $\{ \epsilon^{j,t}_{\nu}, \, O_{\nu},\, a^{(R/I), \nu}_{1m}\}$. Since $a_{1,m}^{\nu}$ in real space indicates the dipole amplitude and direction, they are sampled as global parameters similar to $O_{\nu}$.
The momentum derivative for the dipole coefficients is given by
\begin{equation}\label{P_dot_a1m}
\begin{split}
 \frac{\pd \ln(\calP)}{\pd a^{(R/I), \nu}_{1m}} =
(\pm) 2^{m} \sum_{j} \sum_{i,i' \subset j} Y^{(R/I)}_{1m}(\Omega^j_{i}) &[\Sigma_{\nu}^{-1}]^{jj}_{ii'} [d_{\nu} -  s_{\nu}]^j_{i'},\\
&\text{for $m=0$ and $1$},
\end{split}
\end{equation} 
where $(\pm)$ corresponds to the real ($R$) or imaginary $(I)$ parts respectively, of the coefficients (see \cref{eq:dipole_expr_1}). In general, $s_{\nu}(\Omega^j_{i'})$, is given by \cref{eq:model_multi_t}. Our HMC formalism takes into account pixel-dependent dust emissivity, global offset, and three dipole amplitudes. 

As the algorithm requires, we simulate the Hamiltonian dynamics using the Leap-Frog scheme (see \cref{app:HMC_details}). In the Leap-Frog scheme, the step size $\Delta$ decides the time step, which is generally different for different parameters. A general practice is to standardize the parameter distribution scales, which requires some knowledge of the parameter covariance structure \citep[e.g.][]{2011arXiv1112.4118B}. In the particular case of Gaussian posterior and the model that is linear in parameters, one can choose the mass matrix to achieve this goal. For problems where the curvature is isotropic and constant, such as for the Gaussian likelihood we consider in this work, a parameter independent $\Delta$ can be chosen. This choice of $\Delta$ in the case of Gaussian likelihoods is discussed in \cite{Taylor_Cl_with_HMC_2008}.
For the general distributions with hierarchical modelling or nonlinear parameter dependencies, this procedure may not work as well as it does in our case. By setting the mass matrix equal to the inverse of the covariance matrix of the parameters, $\Delta$ is made independent of the distribution of the individual parameter. The inverse of the parameter covariance matrix is the negative of the parameter Fisher matrix. With our choice of neglecting the correlations between the superpixels, for the given likelihood, the Fisher matrix turns out to be diagonal over the $\epsilon^{j,t}_{\nu}$ parameters. The only non-zero off-diagonal terms are those which connect $\epsilon^{j,t}_{\nu}$ with $O_{\nu}$, $a_{1m}$, and $O_{\nu}$ with $a_{1m}$. However, we neglect these off-diagonal terms. Considering only the diagonal elements while assigning mass for the parameters then does not lead to $\Delta$ for $O_{\nu}$ and $a_{1m}$ being the same as that of $\{ \epsilon^{j,t}_{\nu}\}$. Hence, we choose a different $\Delta$ in the dynamical equations corresponding to $O_{\nu}$ and the dipole parameters. Hence, we have different step sizes in the Leap-Frog scheme, $\Delta_{\epsilon}$ and $\Delta_{O}$ corresponding to $\epsilon^{j,t}_{\nu}$ and $O_{\nu}$, respectively. The details of this choice are discussed in \cref{app:HMC_details}. This leads to the mass matrix with the following terms in the diagonal. The mass for $\epsilon^{j,t}_{\nu}$ is
\begin{equation}\label{eq:mass_emissivity}
    \mu^{j,t}_{\nu} = \sum_{i,i' \subset j} \NHI^{t} (\Omega^j_{i}) [\Sigma_{\nu}^{-1}]^{jj}_{ii'} \NHI^{t}(\Omega^j_{i'}),
\end{equation}
for $O_{\nu}$, it is given by
\begin{equation}\label{eq:mass_offset}
    \mu_{O_{\nu} } = \sum_{j} \sum_{i,i' \subset j} [\Sigma_{\nu}^{-1}]^{jj}_{ii'},
\end{equation}
and the following is the mass for the dipole coefficients
\begin{equation}\label{eq:P2_dot_a102}
\begin{split}
\mu^{(R/I)}_{a_{1m}} = 2^{2m} \sum_{j} \sum_{i,i' \subset j} Y^{(R/I)}_{1m}(\Omega^j_{i})& [\Sigma_{\nu}^{-1}]^{jj}_{ii'} Y^{(R/I)}_{1m}(\Omega^j_{i'}),\\
&\text{for $m=0$ and $1$}.
\end{split}
\end{equation}

Note that the mass matrix elements for $\epsilon^{j,t}_{\nu}$ depend on the noise covariance as well as the templates, whereas those for $O_{\nu}$ and $a_{1m}$ depend only on the noise covariance. In HMC, the proposed parameter is obtained by evolving Hamilton's equations in a certain number of Leap-Frog jumps $N$. The product of $\Delta$ and $N$ determines the total distance traversed in the parameter space and controls the correlation length in the parameter chain. In this section, we have discussed the choices for $N$ and $\Delta$ to simulate the HMC process. While we discussed these choices without much rigour, we tested that the algorithm works using realistic simulations of the data. We would like to point out that formal methods have been developed to tune the HMC algorithm to facilitate appropriate choices for step size and path length.  For example, \cite{2011arXiv1111.4246H} presents the No-U-Turn Sampler scheme to alleviate the need for the user to choose the number of steps and also presents a method for adaptive stepsize. Recent developments in also include SNAPER-HMC for implementation on GPU and TPU hardware \citep{2021arXiv211011576S}, ChEES-HMC \citep{pmlr-v130-hoffman21a}, and various adaptive schemes, for example, MALT-HMC \citep{2022arXiv221012200R}. Some of these schemes are implemented in probabilistic programming frameworks such as STAN \citep{JSSv076i01}, PyMC \citep{2015arXiv150708050S}, and pyro \citep{bingham2019pyro}.

We first validate the above methodology on the Planck simulations. The results obtained from the simulations are presented in \cref{sec:4}. The next section discusses the data, the CIB model, and the sky masks used for our analysis. 

\section{Data Sets} \label{sec:2}
In this section, we describe the Planck data, \hi\ data, and the sky mask used in the analysis. We also describe the procedure for computing the CIB covariance matrix using the model CIB power spectrum.

\subsection{Planck data}\label{sec:2.1}
We use the publicly available Planck 2018 Public Release 3 (PR3\footnote{\url{http://pla.esac.esa.int/pla}}) legacy intensity map at 353\GHz\ \citep{planck-I:2018} for our analysis. The 353 GHz intensity map has been provided in \healpix\footnote{\url{http://healpix.jpl.nasa.gov}} \citep{Gorski:2005} grid at $\Nside=2048$ (pixel size 1.7\arcm) with an angular resolution of FWHM 4.82\arcm\ \citep{planck-III:2018}. We subtract the CMB contribution at 353 GHz using the following procedure. We use the \smica\ CMB map provided at a beam resolution of 5\arcm\ (FWHM) and $\Nside = 2048$ \citep{planck-IV:2018}. We smooth the 353 GHz intensity map at the resolution of the \smica\ CMB map using the Gaussian approximation of the Planck beam and subtract the contribution of CMB. We further smooth the CMB-subtracted Planck 353 GHz map at the beam resolution of 16.2\arcm\ (FWHM), downgrade to $\Nside = 512$ (pixel size 6.8\arcm) resolution. We choose the Planck 353 \GHz\ map for our analysis because the contribution of synchrotron and free-free emissions are negligible compared to that of dust after the contribution due to the CMB is subtracted. We consider the Planck CMB-subtracted intensity map as the primary data. We treat the CIB monopole term as the global offset parameter. We use the unit conversion factors mentioned in \cite{planck-IX:2014} to convert 353 \GHz\ map from \kcmb\ to \KJysr\ unit.

We use 300 end-to-end (E2E) noise realizations for the Planck frequency maps from the Planck Legacy Archive \citep{planck-XI:2018}. The original maps are provided at $\Nside= 2048$. We smoothed all the 300 noise maps at 16.2\arcm\ FWHM beam resolution and re-project them at the resolution of $\Nside=512$. Then, we compute the variance map using the 300 smoothed E2E noise maps. 


\subsection{\hi\ data}\label{sec:2.4}
To separate the \hi-correlated dust emission from the Planck data at high Galactic latitude, we use the \hi\ data provided by  GASS\footnote{\url{https://www.atnf.csiro.au/research/GASS/Data.html}} survey carried out by Parkes telescope \citep{GASSI:2009}. The survey observed 21 cm emission over the southern galactic sky (at declinations $\delta < 1\deg$) within velocity range, $-400 \kms < V_{\rm LSR} < 500 \kms$; where $V_{\rm LSR}$ is the velocity of the \hi\ clouds with respect to local standard of rest. The survey has a beam resolution of 14.5\arcm\ FWHM, velocity resolution $\delta v = 1 \kms $, and root-mean-square brightness temperature uncertainty of 50 mK ($1\sigma$). The GASS survey maps used in our analysis are from \cite{Kalberla:2010} and are corrected for instrumental effects, stray radiation, and radio-frequency interference. 

In the southern Galactic cap, the \hi\ in the Galactic disk (or what we call Galactic \hi) is mixed with significant emission from the Magellanic Stream (MS) (\citealt{Nidever:2008, Elena:2016}). The spectra in the 3D data cube (longitude, latitude and radial velocity) likely to be associated with MS are distinguished from Galactic \hi\ spectra using the velocity information \citep{Venzmer:2012}. The three-dimensional model of \cite{Kalberla:2008} helps to distinguish the spectra associated with the Galactic \hi\ emission and MS. Finally, the \hi\ template map is produced integrating 3D spectra over velocity range and projected on \healpix\ grid at $\Nside=1024$. The Galactic \hi\ column density map \NHI\ used in our analysis has an angular resolution of 16.2\arcm\ FWHM and is projected on the \healpix\ grid $\Nside=512$ (pixel size 6.9\arcm). We use the Galactic \hi\ map as a tracer for the \hi-correlated dust emission. The same Galactic \hi\ map is used by the Planck collaboration to study dust emission properties in the diffuse interstellar medium \citep{planck-XVII:2014}.

\subsection{Sky masking} \label{sec:masking}
We use the same mask as used in \cite{planck-XVII:2014} for our dust-\hi\ correlation analysis. The total sky area of the mask is 6300\,deg$^2$ ($f_{\rm sky} = 15.3\%$) where $\NHI  < 6 \times 10^{20} \cm2$, and thus avoids the high column density regions. The unmasked sky region covers the Galactic latitude $b \leq -25\deg$. The area of 20\deg\ diameter centred around ($l_{MS}$, $b_{MS}$) = ($-50\deg$, 0$\deg$) is masked to avoid Magellanic Stream \citep{Nidever:2010}. Further, the bright radio sources at microwave frequencies and infrared galaxies at 100 \mum\  have been masked out. 

To test the dependence of the analysis results on the sky fraction, we generate four additional masks with different \NHI\ cutoff over the range between $2\times10^{20} \text{cm}^{-2}$ and $5 \times10^{20} \text{cm}^{-2}$. We mask the regions with \NHI\ values higher than the cutoff value. We label the mask with \NHI\ cutoff $Q \times 10^{20} \text{cm}^{-2}$ as MaskHIQ. The sky masks are overlapping by construction. We use overlapping masks to study the dependence of the global offset as a function of mean \hi\ column density. The respective sky fraction $f_{\rm sky}$ for each sky mask is quoted in \cref{table:offset}.

\section{Planck simulations}\label{sec:4}

In this section, we validate our methodology on the Planck simulations to simultaneously fit a dust emissivity per superpixel and a global offset using the HMC sampler. 

We simulate the dust intensity maps at Planck HFI frequency bands between 217 and 857 GHz. We analyse simulated maps considering the total noise contribution from the instrument noise and the CIB anisotropies. We ignore the contribution of Galactic residuals in this analysis. We consider the instrument noise to be uncorrelated between pixels. However, for the CIB, we consider the inter-pixel correlations. Analysis with the simulated data helps to validate the pipeline and to determine some analysis choices, for example, the optimal values of the Leap-Frog step size $(\Delta)$ and the Leap-Frog jumps ($N$) by examining the behaviour of the Markov Chains (MC) drawn from the posterior distribution. 

We require sufficient unmasked subpixels within each superpixel to fit the signal model with the data. We set this threshold to one-third of the subpixels within each superpixel\footnote{Here, the threshold is 85 unmasked subpixels within superpixels at $\Nside = 32$.}. We excluded those superpixels from the joint fitting which do not fulfil this criterion.

\subsection{Simulated maps}\label{sec:5.1}
We construct the simulated maps using the following procedure:
\begin{enumerate}

\item We start with the dust emissivity map of \cite{planck-XVII:2014} at 353 GHz ($\epsilon_{353}$) projected on $\Nside=32$ (low-resolution) \healpix\ grid. This map is obtained through the dust-\hi\ correlation analysis over 15\deg\ circular patches in diameter centred on \healpix\ pixels at $\Nside=32$.
We assume that $\epsilon_{353}$ is the same at all subpixels defined at resolution $\Nside=512$ that fall within a superpixel defined by $\Nside=32$. Each superpixel contains 256 subpixels. 

\item We translate the dust emissivity map from 353\, GHz to other HFI frequencies using the MBB spectrum,
\begin{equation}
\epsilon_{\nu} (\Omega^j_{i}) = \epsilon_{353} (\Omega^j_{i})\left(\frac{\nu}{353}\right)^{\beta_{\rm d}}\frac{B_{\nu}({T_{\rm d}})}{B_{353}({T_{\rm d}})}, 
\label{eq:4.1}
\end{equation}
where $\beta_{\rm d}$ is the dust spectral index fixed to 1.5, $B_{\nu}$ is Planck black-body function and $T_{\rm d}$ is the dust temperature fixed to 20\,K \citep{Planck_PIPXXII:2015}. 

\item We simulate the dust intensity maps at 217, 353, 545, and 857 GHz at $\Nside= 512$ using the dust emissivity map and Galactic \hi\ template. We add the global offset values from \cite{planck-XLVIII:2016} to produce the signal model maps at all HFI frequencies. 

\item To simulate the instrument noise contribution, we use the variance map ($II$) computed from 300 smoothed E2E noise maps. We assume the instrumental noise to be Gaussian, white, and uncorrelated between the pixels. For the CIB noise component, we simulate the CIB map smoothed at 16.2\arcm\ FWHM beam resolution projected at $\Nside=512$ from the model CIB power spectrum at each HFI frequency. Though the CIB is correlated between two frequencies, analysing individual frequency maps entails neglecting the correlation between the CIB emission at different frequencies. 

\item Finally, we co-add simulated dust intensity, global offset, instrument noise, and the CIB, all expressed in \KJysr\ units. 

\item To build the likelihood, we construct the instrumental noise covariance matrix ($\Sigma^{N}_{\nu}$) and the CIB covariance matrix ($\Sigma^{\rm CIB}_{\nu}$). $\Sigma^{N}_{\nu}$ is taken as a diagonal covariance matrix because we neglect the inter-pixel instrument noise. $I^{\rm CIB}_{\nu}$ exhibits a significant correlation between subpixels within a given superpixel. 
\end{enumerate}
\begin{figure}
\centering
\includegraphics[width=\columnwidth]{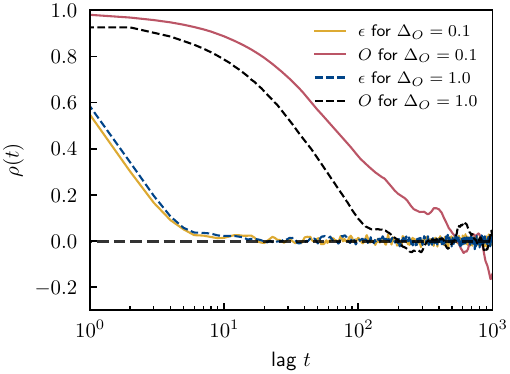}
\caption{The auto-correlation coefficient, $\rho(t)$, at 353 GHz for the samples of the emissivity in one superpixel ($\epsilon$) and the global offset ($O$) as a function of the lag $(t)$ for two different choices of Leap-Frog step size ($\Delta_{O}$) for the global offset and fixed $\Delta_{\epsilon} = 0.1$. For $\Delta_{O} = 0.1$, the correlation length for $\epsilon$ and $O$ are $8$ and $220$, respectively. For $\Delta_{O} = 1.0$, the correlation length for $\epsilon$ in the same pixel and $O$ are $10$ and $102$, respectively.}
\label{fig:4_sec4}
\end{figure}
\begin{table*}
\caption{Results of analysis of simulated maps at four Planck HFI frequencies. The global offset values with corresponding 1$\sigma$ error bars are estimated over five sky masks, tracing low to intermediate \hi\ column density regions. The inference of offset is stable with respect to sky coverage, with a slight decrease in the uncertainty with higher $f_{\rm sky}$. }
\label{table:offset}
\begin{tabular}{|c|c|ccccc|}
\hline
Frequency & Input offsets & \multicolumn{5}{c|}{Recovered Offsets [\KJysr]}  \\
$[\GHz]$ & [\KJysr]  & \multicolumn{5}{c|}{ }  \\
\cline{3-7}
  & &   \multicolumn{5}{c|}{Sky masks}  \\ 
& & MaskHI2 & MaskHI3 & MaskHI4  & MaskHI5  & MaskHI6 \\
\cline{3-7}
& & \multicolumn{5}{c|}{$f_{\rm sky}$ [\%]}  \\ 
& &  7.3& 11.5 & 13.9 & 15.0 & 15.3  \\ 
\hline
217 & 40&\hspace{1.3mm}$40.0 \pm 0.1$ & \hspace{1.3mm}$40.0 \pm 0.1$  & \hspace{1.3mm}$40.0 \pm 0.1$  & \hspace{1.3mm}$40.0 \pm 0.1$ & \hspace{1.3mm}$40.0 \pm 0.1$ \\
353 & 120&$119.6 \pm 0.4$ & $120.1 \pm 0.4$  & $119.9 \pm 0.3$  & $119.9 \pm 0.3$ & $120.0 \pm 0.2$ \\
545 & 330&$330.9 \pm 0.9$ & $330.5 \pm 0.7$  & $330.3 \pm 0.7$  & $330.6 \pm 0.7$ & $330.6 \pm 0.6$ \\
857 & 550&$550.5 \pm 1.2$ & $550.6\pm 1.0$  & $550.6 \pm 0.9$  & $550.7 \pm 0.9$ & $550.6 \pm 0.9$ \\
\hline 
\end{tabular}
\end{table*}
 
\subsection{Validation with simulations}\label{sec:5.3}
We focus our discussion of the simulated map analysis on the 353 GHz frequency channel without the loss of generality. However, we summarise the simulation results at all four HFI frequencies ($217-857$ GHz).

The output of our HMC algorithm is the chains of MC samples of the dust emissivity per superpixel at $\Nside=32$ and the global offset. The total number of parameters at each frequency is 2011 (dust emissivity values) + 1 (global offset) over MaskHI6. 
\begin{figure}
\centering
\includegraphics[width=\columnwidth]{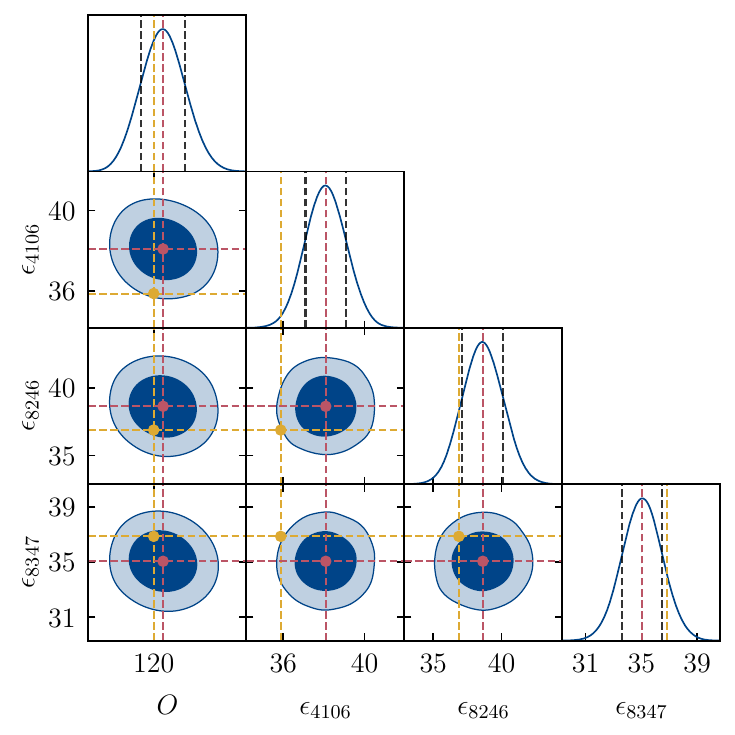} 
\caption{The joint and marginalised probability distributions of dust emissivities (expressed in $\KJysr (10^{20}\,\text{cm}^{-2})^{-1}$) at three representative superpixels (pixel index as super-script) and the global offset (in $\KJysr$) at 353 GHz. The red lines mark the posterior mean, and the orange lines depict the input values of the respective parameters. The contours mark the $68\%$ and $90\%$ regions of the joint distributions. The vertical black dashed lines in the histogram mark $16$, $50$, and $84$ percentiles of the distribution.}
\label{fig:5_sec4}
\end{figure}

We obtain $2 \times 10^{4}$ samples for each derived parameter and discard the first $10^{3}$ samples. We use the remaining $1.9 \times 10^{4}$ MC samples for further analysis. In practice, the samples for a given parameter are not independent but are correlated with a certain correlation length. \cref{fig:4_sec4} presents the auto-correlation coefficients, $\rho(t)$, for sample chains of the dust emissivity at one superpixel and the global offset chains for 353 GHz maps. The results are depicted for two choices of $\{ \Delta_{\epsilon}, \Delta_{O} \}$. When step size for $\epsilon$ and ${O}$ are the same, $\{ \Delta_{\epsilon}, \Delta_{O} \} = \{ 0.1, 0.1\}$, the global offset (magenta solid) chain is highly correlated. For this choice, the correlation length of emissivity and offset chain is around 8 and 220, respectively. For $\{ \Delta_{\epsilon}, \Delta_{O} \} = \{ 0.1, 1.0\}$, the global offset (red dashed) chain becomes less correlated, and correlation length decreases by a factor of 2. However, the correlation length for the emissivity chain remains almost the same. For the latter choice of step sizes, the correlation length of dust emissivity and global offset chains are around 10 and 102, respectively. Therefore, we adopt the second choice of step sizes for the HMC sampling at all frequencies. Along with these $\{\Delta_{\epsilon}, \Delta_{O}\}$, we choose the number of Leap-Frog steps $N = 10$, which gives a reasonable acceptance rate and lower correlation length. We check the convergence of the chains using the Gelman-Rubin test \citep{Gelman-Rubin:1992} (for details, see \cref{sec:5.2}). Using seven independent chains, we find Gelman-Rubin Markov Chain Monte Carlo (MCMC) convergence diagnostics are 1.0002 (for emissivity) and 1.002 (for offset). These values confirm the chains are converged to a reasonable accuracy.
\begin{figure}
\includegraphics[width=\columnwidth]{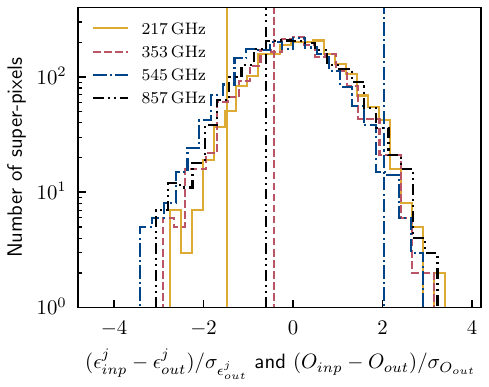}
\caption{The histograms of the normalised difference of input emissivities and the posterior mean emissivities, $\delta^j=(\epsilon^j_{inp}-\epsilon^j_{out})/\sigma_{\epsilon^j_{out}}$ for four Planck frequencies from 217 to 857 GHz. Vertical lines depict the normalised difference between the input and output global offset at respective Planck frequencies. }
\label{fig:6_sec4}
\end{figure}

\begin{figure}
\includegraphics[width=\columnwidth]{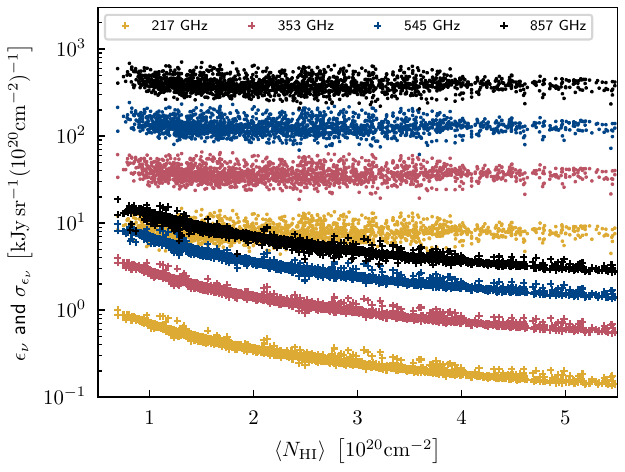} 
\caption{The recovered dust emissivity $\epsilon^j_{\nu}$ and its standard deviation $\sigma_{\epsilon^j_{\nu}}$ per superpixel ($j$) as a function of mean Galactic \hi\ column density ($\langle \NHI (\Omega^j_{i}) \rangle$). The mean $\langle..\rangle$ is taken over all unmasked subpixels ($i$) that fall within a given superpixel. The ``dot" symbol represents $\epsilon^j_{\nu}$ and ``plus" represents $\sigma_{\epsilon^j_{\nu}}$ at all Planck frequencies considered in this analysis.}
\label{fig:em_vs_nhi}
\end{figure}

\begin{figure*}
\centering
\includegraphics[width=\linewidth]{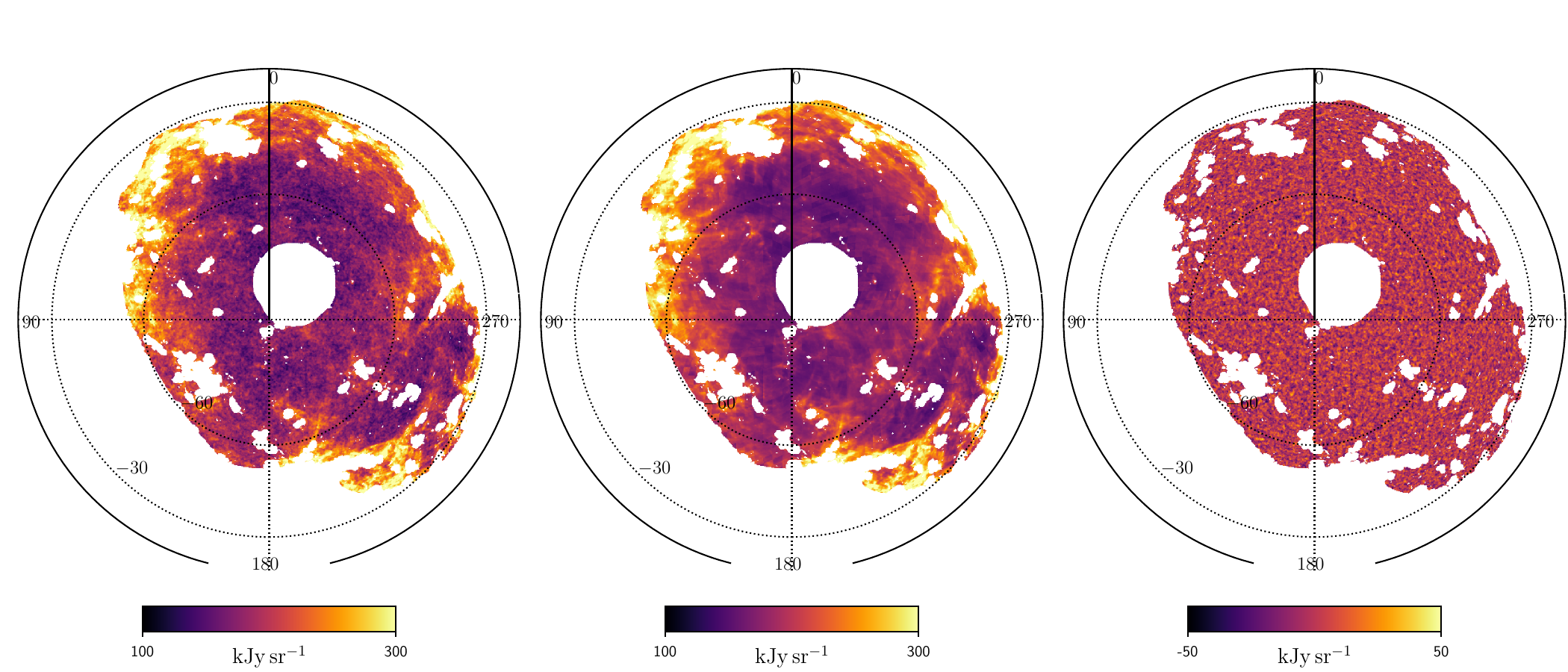}
\caption{The left panel shows the input simulated map at 353 GHz (in units of \KJysr) over MaskHI6. The middle panel shows the signal model map (in \KJysr) derived from the mean values of the emissivity and the offset obtained using the HMC sampler. The right panel depicts the residual map (in \KJysr) obtained by subtracting the signal model map from the input map at 353 GHz. The residual map has contributions from the CIB and the instrumental noise. }
\label{fig:7_sec4}
\end{figure*}
\begin{figure*}
\includegraphics[width=\linewidth]{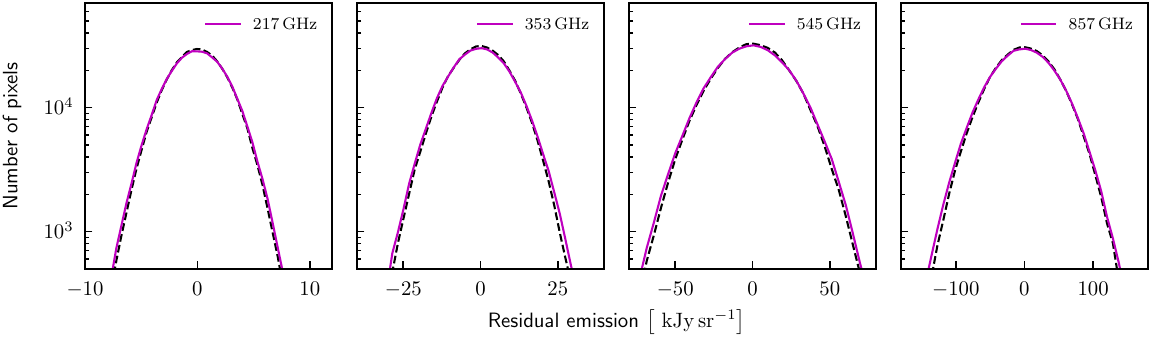} 
\caption{Figure shows the histograms of the residual map (black curve) after subtracting the signal model from the input simulated map at Planck HFI frequencies over the unmasked pixels of eMaskHI6. The expected histograms of the residuals (magenta curve) are produced from a single realization of instrument noise and the CIB at the respective frequency over the same sky mask. The agreement between the observed and expected residual distribution validates an unbiased inference of emissivity and the offset for realistic Planck simulations.}
\label{fig:8_sec4}
\end{figure*}

To check the correlations between the model parameters, we show the joint distributions of the dust emissivity at three superpixels and the global offset in \cref{fig:5_sec4} at 353 GHz. We do not find a significant correlation between the dust emissivities at two different superpixels or between the emissivity at a given superpixel and the offset. We also show the marginalised posterior distributions of the respective parameters along with their posterior mean values (red dashed line) and corresponding input values (orange dashed line) used in the Planck simulation. We find the inferred posterior mean values of the parameters agree with their respective input values.
\begin{figure}
\includegraphics[width=\columnwidth]{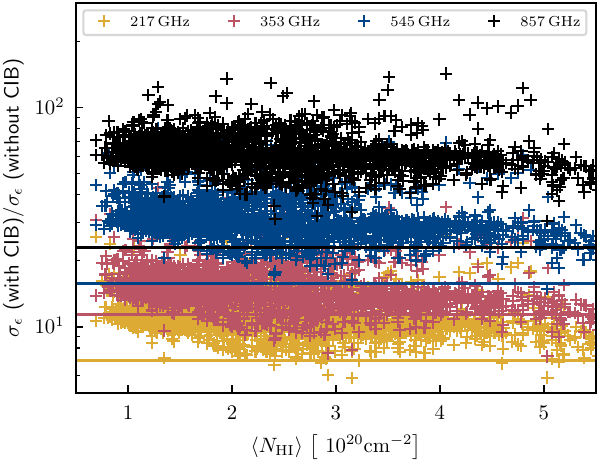} 
\caption{The ratio of $1\sigma$ uncertainty in dust emissivity with and without the CIB contribution in the noise covariance term. The ratio is plotted for all the superpixels over the unmasked sky area against their mean \NHI\ value (``plus" marker). The horizontal lines represent the ratio of the uncertainty in the offset parameter, which is a single number at a given frequency, with and without the CIB in the noise covariance.}
\label{fig:9_sec4}
\end{figure}

To quantify the goodness of fit, we use posterior predictive checks considering the pixels included in the analysis, which are a subset of those in the MaskHIQ mask. This is due to the exclusion of superpixels that do not satisfy the threshold criteria for the number of subpixels. Most of the boundary pixels in the MaskHIQ mask do not satisfy the criteria and are excluded from further analysis. We term the resultant extended mask as eMaskHIQ. We replicate the simulated data based on the mean and standard deviation of the parameter samples. The distribution of the original simulated data and the replicated data match very well. Both the distributions are non-Gaussian due to the non-Gaussian nature of the thermal dust emission, which has a dominant contribution. We use summary statistics like median and 95\% quantile level to test the consistency between the original and the replicated simulated data. The probability-to-exceed (PTE) statistics are used to test the probability that the summary statistics of the replicated data exceed the original data. The PTE values of the summary statistics are quoted in Table~\ref{table:sim-post-test}.

\begin{table}
\begin{center}
\caption{The PTE values obtained for all Planck frequencies considered for our analysis over eMaskHI6. They are defined as the probability of obtaining summary statistics (median and 95\% quantile) larger than fitted data, based on 1000 simulations with dust plus CIB and Planck noise. The PTE values are expressed as a percentage.}
\label{table:sim-post-test}
\begin{tabular} { |c|  c c| }
\hline
 \thead{Frequency \\$[\GHz]$} &  Median ($\%$) &  $95\%$ quantile ($\%$)\\
\hline
$217$ & $11$ & $34$\\
$353$ & $63$ & $48$\\
$545$ & $98$ & $96$\\
$857$ & $55$ & $70$\\
\hline
\end{tabular}
\end{center}
\end{table}

From the MC chains of the parameters, we compute the mean and standard deviation of the respective parameters. In order to quantify the accuracy of inference, in \cref{fig:6_sec4},  we present the distribution of the standard deviation 
($\sigma_{\epsilon^j_{out}}$) normalised difference, $\delta^j = (\epsilon^j_{inp} - \epsilon^j_{out})/\sigma_{\epsilon^j_{out}}$ of input dust emissivity at all superpixels ($\epsilon^j_{inp}$) and the posterior mean emissivity at respective superpixels ($\epsilon^j_{out}$), for all four frequency bands. We also show the same quantity for the offset with vertical dashed lines. The mean values obtained using the MC samples agree very well with their respective mean values and are within $3\sigma$ deviations. The largely symmetric nature of the histograms implies that the best-fit values of output dust emissivities for all superpixels are unbiased.

In \cref{fig:em_vs_nhi}, we show the mean and standard deviation of dust emissivity per superpixel obtained from the dust-\hi\ correlation analysis as a function of mean Galactic \hi\ column density at those superpixels. The uncertainty decreases with an increase in mean Galactic \hi\ column density, indicating a better estimation of dust emissivity in high column density regions within the sky mask MaskHI6. This is expected as the uncertainty scales with $1/\mu_{O_{\nu}}$ (see \cref{eq:mass_offset}) and is inversely proportional to \NHI. 

We use the mean of the parameter samples as the best-fit value of the respective parameter \citep{Mackay:2003}. Owing to the linear nature of the signal model, the signal that corresponds to the best-fit values of emissivity and the offset is also the best-fit signal. We obtain the best-fit intensity map corresponding to the signal model using the mean values of the dust emissivity and the global offset following \cref{eq:model}. In \cref{fig:7_sec4}, we show the simulated map at 353 \GHz\ along with maps of the best-fit model and the residual. The residual map is the difference between the input and best-fit model intensity maps. The residual map contains the contribution from the instrument noise and the CIB. The residual map shows the small-scale structure, lacking large-scale fluctuations, consistent with the nature of instrument noise and the CIB. While the map shows the spatial distribution of the residual, to check the nature of the distribution of the residual, in \cref{fig:8_sec4}, we compare it (black)  with the expected residual contribution from instrument noise and the CIB (magenta). We find good agreement between the recovered and the expected residual distribution at all Planck frequencies, indicating that the analysis does not introduce systematic bias. At the map level, the residual map over the unmasked region strongly correlates with the expected residual map obtained by combining the input CIB and instrumental noise map at all Planck frequencies. The Pearson correlation coefficients are 0.97, 0.96, 0.96, and 0.96 at 217, 353, 545, and 857 GHz, respectively.

Offset is a pixel-independent parameter; hence, we expect its inference to be unaffected by mask choice. We infer the global offset values for different mask choices using simulated maps to test this. In \cref{table:offset}, we list the posterior mean values of the global offsets along with 1$\sigma$ error bars and the respective masks. Irrespective of the choice of mask, the inferred offset values at all frequencies are consistent with their input values. This indicates the stability of the analysis for different choices of masks. 

To elucidate the effect of the CIB noise in our analysis, we redo the analysis without the CIB contribution in the noise covariance matrix. In \cref{fig:9_sec4}, we compare the standard deviation of the dust emissivity and offset inferred with and without considering the CIB in the noise covariance matrix. We plot the ratio of these two standard deviations for all the superpixels against the mean $\langle \NHI \rangle$ value at the respective superpixel. The uncertainty ratio in the offset parameter with and without the CIB for all four frequencies is depicted with horizontal solid lines. There is \ a weak dependence between the ratio of $\sigma_{\epsilon}$ with and without the CIB as a function of \NHI\ value. The ratio increases with the increasing frequency, consistent with the fact that the CIB noise dominates over instrumental noise at higher HFI frequency \citep{planck-XVII:2014}. 

We have shown that the method gives an unbiased inference of emissivity and the offset in the presence of realistic noise. We can faithfully take into account the noise arising due to the CIB, including the CIB-induced inter-pixel correlations within a superpixel, while we neglect the correlation between subpixels belonging to two different superpixels. The implication of considering the CIB noise is evident in \cref{fig:9_sec4}. Not considering the CIB can lead to underestimating the uncertainty by orders of magnitude and possible bias in the inference. Further, joint sampling of emissivities with the global offset mitigates any bias that may arise due to the biased or position-dependent value of the offset.

In this simulation section, we completely ignore the contribution of the Galactic residual. Assuming the same SED for \hi-correlated dust emission and Galactic residuals, one can expect it to dominate over the CIB at Planck's highest frequency, 857 GHz. Like the CIB emission, we can incorporate the contribution of Galactic residuals in the noise covariance term.

\subsection{Validation of two-template fit with Galactic \hi\ and MS templates}\label{sec:two_temp-fit}
\begin{figure}
\includegraphics[width=\columnwidth]{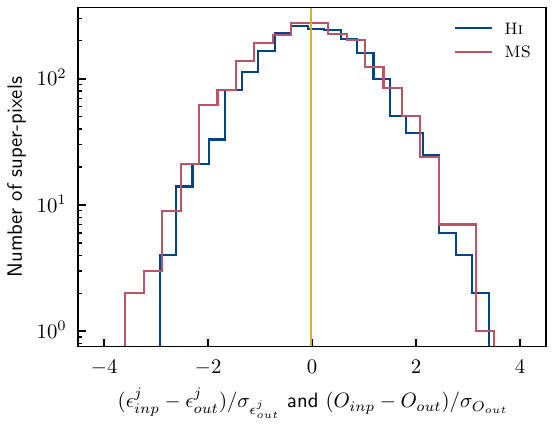}
\caption{The histograms of the normalised difference of input and the posterior mean emissivities for Galactic \hi\ and MS template at 353 GHz. The vertical line shows the normalised difference between the input and output global offset at the same frequency.}
\label{fig:two_template_fit}
\end{figure}

To test the method's robustness, we repeat the same analysis on the 353 GHz simulated map that has a contribution from the two \NHI\ templates. We use the MS and Galactic \hi\ templates to simulate map 353 GHz. The MS template traces the gas from IVC and HVC. We add the MS template with a constant emissivity $\epsilon^{\mathrm{MS}}_{353} = 10^{-2}\langle \epsilon^{\hi}_{353}\rangle$, where $\langle \epsilon^{\hi}_{353}\rangle$ is the average over all the superpixels of the input dust emissivity map ($\epsilon_{353}$ map as used in \cref{sec:5.1}). The corresponding input values for the dust emissivities are $\langle \epsilon^{\hi}_{353}\rangle = 37.3\,\KJysr (10^{20} \mathrm{cm^{-2}})^{-1}$ and $\epsilon^{\mathrm{MS}}_{353} = 0.37\KJysr (10^{20} \mathrm{cm^{-2}})^{-1}$, respectively, and the offset is $120\KJysr$, the same as that used in \cref{sec:5.1}. We infer the global offset and the emissivity parameter per superpixels for both the Galactic \hi\ template and the MS template over MaskHI6 performing the HMC methodology as discussed in \cref{sect:2.2}. \cref{fig:two_template_fit} shows that the recovered emissivity values are well within $3\sigma$ of the input value without significant bias. The inferred global offset is $120.0 \pm 0.3\, \KJysr$, which is close to the input global offset value. The recovered emissivities are quoted as the mean of the dust emissivity values over all valid superpixels, along with its standard deviation. They are respectively, $37.3\,\KJysr (10^{20} \mathrm{cm^{-2}})^{-1}$ and $7.2\,\KJysr (10^{20} \mathrm{cm^{-2}})^{-1}$ for the Galactic \hi\ template while $0.39\,\KJysr (10^{20} \mathrm{cm^{-2}})^{-1}$ and $0.27\,\KJysr (10^{20} \mathrm{cm^{-2}})^{-1}$ for the MS template.

\subsection{Validation with a residual CMB dipole term}\label{sect:dipole_sim}
\begin{table*}
\caption{Recovered global offset ($O$), dipole amplitude ($A_{\mathrm{dip}}$) and Galactic longitude ($l_{\mathrm{dip}}$) and latitude ($b_{\mathrm{dip}}$) estimated over five different \NHI\ cutoff sky masks for simulated map at 217 GHz. The corresponding input values are respectively $O = 40\, \KJysr, A_{\mathrm{dip}}=3.9\, \KJysr, l_{\mathrm{dip}}=264.0\degree, b_{\mathrm{dip}}=48.3\degree$. Column $\delta$ represents the difference between the input and the recovered parameter value in units of $1\sigma$ uncertainty.}
\label{table:dipole-mask}
\begin{tabular} { |c|  c c c c c c c c|}
\hline
Sky masks &  \multicolumn{2}{c}{\thead{$O\,\left[\mathrm{kJy\,sr^{-1}}\right]$}} &  \multicolumn{2}{c}{$A_{\mathrm{dip}}\,\left[\mathrm{kJy\,sr^{-1}}\right]$} &  \multicolumn{2}{c}{$l_{\mathrm{dip}}\,\left[\mathrm{degree}\right]$} &  \multicolumn{2}{c|}{$b_{\mathrm{dip}}\left[\mathrm{degree}\right]$}\\
\cline{2-9}
  &  \thead{Recovered} &  $\delta$ &  Recovered &  $\delta$ &  Recovered &  $\delta$ &  Recovered &  $\delta$\\
\hline 
MaskHI2 & $39.7\pm1.6$ & $0.2$ & $4.1\pm0.8$ & $-0.3$ & $287.2\pm9.4$ & $-2.5$ & $38.4\pm24.6$ & $0.4$\\
MaskHI3 & $39.3\pm0.9$ & $0.7$ & $3.6\pm0.5$ & $0.6$ & $271.2\pm6.5$ & $-1.1$ & $34.0\pm15.9$ & $0.9$\\
MaskHI4 & $39.2\pm0.7$ & $1.1$ & $3.8\pm0.4$ & $0.2$ & $273.0\pm4.2$ & $-2.1$ & $30.5\pm12.5$ & $1.4$\\
MaskHI5 & $39.8\pm0.6$ & $0.4$ & $3.9\pm0.4$ & $-0.1$ & $273.1\pm4.7$ & $-1.9$ & $39.7\pm10.0$ & $0.9$\\
MaskHI6 & $39.2\pm0.6$ & $1.2$ & $3.7\pm0.3$ & $0.6$ & $271.4\pm4.2$ & $-1.7$ & $31.0\pm10.9$ & $1.6$\\
\hline
\end{tabular}
\end{table*}

Given that we infer a global offset parameter from the partial sky, the presence of residual CMB dipole can bias its inference. We demonstrate that the method can infer a residual CMB dipole and the global offset jointly. We use this exercise to assess the extent to which the residual dipole affects the inference of the global offset from the partial sky. 

\begin{figure}
    \centering
    \includegraphics[width=\columnwidth]{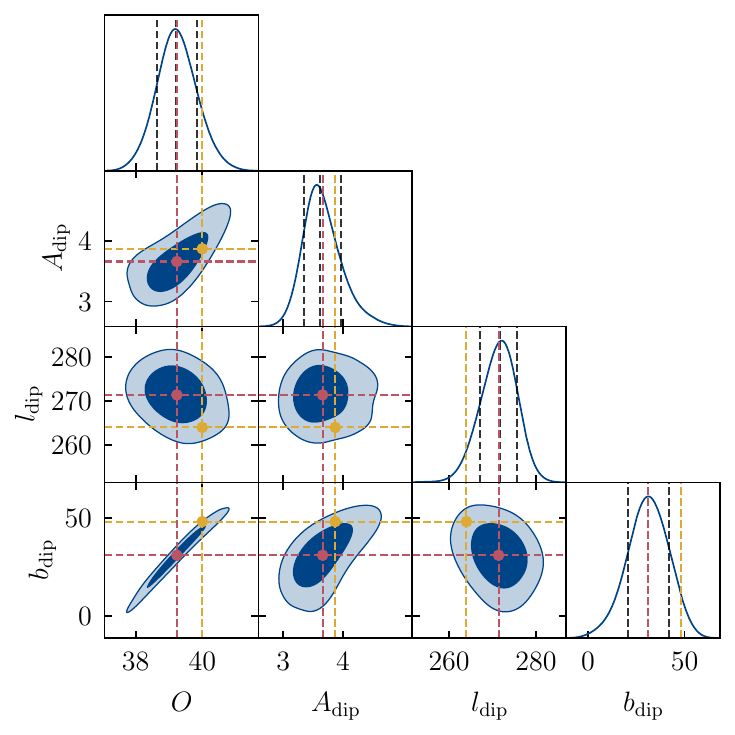}
    \caption{The joint and marginalised probability distributions of global offset and the three dipole parameters at 217 GHz. The global offset and residual CMB dipole amplitude are expressed in $\KJysr$ unit and the residual CMB dipole direction in degrees. Details are the same as Figure \ref{fig:5_sec4}.}
    \label{fig:dipole_triangle}
\end{figure}

We simulate a single realization of the Planck map at 217 GHz with the residual CMB dipole amplitude $A_{\mathrm{dip}} = 3.9 \KJysr$, corresponding to $8\,\mu \mathrm{K}$ in \kcmb\ unit. We choose this representative amplitude approximately equal to the difference between WMAP and Planck estimates of the Solar system dipole \citep{planck-I:2018}. The direction $\left(l_{\mathrm{dip}}, b_{\mathrm{dip}}\right) = \left(264.0\degree, 48.3\degree\right)$ is chosen same as the Solar dipole \citep{2014A&A...571A..27P}. Corresponding to these real space dipole coordinates, the input values of the harmonic coefficients in \cref{eq:dipole_expr_1} are $(a_{1,0}, a^{R}_{1,1}, a^{I}_{1,1}) = (5.9, 0.4, -3.7)$ \KJysr. The value of the global offset is $O = 40\,\KJysr$ (same as in \cref{table:offset}), and the input dust emissivity averaged over the superpixels for 217 GHz is $8.15\KJysr (10^{20} \mathrm{cm^{-2}})^{-1}$. Following the procedure detailed in \cref{sect:2.2}, we jointly sample emissivity per superpixel, the global offset, and the harmonic coefficients using variable Leap-Frog jumps to reduce the correlation among the parameters (as discussed in \cref{app:var-nsteps}). We recover the offset $O = 39.2 \pm 0.6 \KJysr$ and the harmonic coefficients as $a_{1,0}= 3.9 \pm 1.5$, $a^{R}_{1,1} = -0.1\pm0.3$ and $a^{I}_{1,1} = -4.4\pm0.4$ in $\KJysr$ for maskHI6. The recovered emissivity is within $3\sigma$ as verified from a normalised histogram, indicating the method works well. The harmonic coefficients are related to the dipole amplitude and Galactic longitude and latitude, respectively, as
\begin{align}
A_{\mathrm{dip}} &= \sqrt{\frac{3}{4\pi}} \sqrt{a_{1,0} + 2 a^{R}_{1,1} + 2 a^{I}_{1,1}},\nonumber\\
l_{\mathrm{dip}} &= \arctan\left(-\frac{a^{I}_{1,1}}{a^{R}_{1,1}}\right),\label{eq:dipole_alm_to_amp}\\
b_{\mathrm{dip}} &= \frac{\pi}{2} - \arccos\left(\frac{a_{1,0}}{A_{\mathrm{dip}}}\sqrt{\frac{3}{4\pi}}\right).\nonumber
\end{align}
We calculate $A_{\mathrm{dip}}$ and $(l_{\mathrm{dip}}, b_{\mathrm{dip}})$ for each $1.9\times 10^{4}$ samples using  \cref{eq:dipole_alm_to_amp}. The joint and marginalised probability distributions of the global offset, dipole amplitude and direction are shown in Figure \ref{fig:dipole_triangle}. The corresponding mean values with standard deviations are obtained as $A_{\mathrm{dip}} = 3.7\pm 0.3\,\KJysr$ and direction $\left(l_{\mathrm{dip}}, b_{\mathrm{dip}}\right) = \left(271.4\pm4.2 \degree, 31.0\pm10.9\degree\right)$. 

We find that the amplitude of the fitted dipole is positively correlated with the global offset. This is expected due to the dipole direction lying in the northern Galactic part, which is almost opposite to the SGP. The increased dipole amplitude corresponds to more negative dipole fluctuation in the SGP region, which is compensated by the global offset parameter increase. A similar reason leads to the positive correlation between the $b_{\mathrm{dip}}$ and the offset. Lower values of $b_{\mathrm{dip}}$ correspond to the most negative part of the dipole pointing away from the SGP, leading to the dipole compensating for a greater fraction of the positive zero level and letting the offset have a relatively smaller value. Including dipole in the analysis does not significantly change the inference of the offset, indicating unbiased inference. However, due to the correlated nature of the offset and the dipole parameters, there is a significant increase in the uncertainty of the offset parameter. \cref{table:dipole-mask} presents the recovered offset and dipole parameters as a function of sky masks at 217 GHz. As the sky area increases, the uncertainty of the global offset and three dipole parameters decreases. 

We repeat the analysis with the residual CMB dipole at 353 GHz using the same Planck simulations. The same residual CMB dipole amplitude at 353 GHz is $A_{\mathrm{dip}} = 2.4\,\KJysr$ (in intensity units). We performed the same analysis and recovered the offset $O=119.0\pm 2.4\,\KJysr$ and dipole amplitude $A_{\mathrm{dip}} =3.7 \pm 1.4\,\KJysr$ and direction as $\left(l_{\mathrm{dip}}, b_{\mathrm{dip}}\right) = \left(277.4\pm 31.1\degree, 14.9\pm 43.2\degree\right)$. The difference between input and the recovered values of the $O$, $A_{\mathrm{dip}}$, $l_{\mathrm{dip}}$ and $b_{\mathrm{dip}}$ are respectively $0.4\sigma$, $-0.9\sigma$, $-0.4\sigma$ and $0.8\sigma$, $\sigma$ being the uncertainty on the respective parameters. The uncertainty of inferred dipole parameters at 353 GHz is larger than that at 217 GHz. This is due to higher noise at 353 GHz, where the contribution from the CIB is higher than at 217 GHz (see \cref{fig:2}). The lower amplitude of residual CMB dipole at 353 GHz (in intensity units) and higher uncertainty results in a low signal-to-noise ratio, but the recovered value is consistent with the input within $1\sigma$. Our results show that the residual CMB dipole contribution can be ignored at frequencies 353 GHz and above. We conclude that at the noise level considered here, if the offset is larger than the amplitude of the dipole, neglecting the residual CMB dipole would not lead to significant bias in the inference of global offset. 

\section{Planck data results}\label{sec:5}
In this section, we discuss the analysis results of the 353 GHz CMB-subtracted Planck intensity map. We model the data with pixel-dependent dust emissivity at $\Nside=32$ resolution and a global offset. For this analysis, we do not consider the residual CMB dipole contribution in the signal model. As shown with the simulations, the noise at 353 GHz leads to increased uncertainty on the residual CMB dipole parameters as compared to the same inference at 217 GHz. However, we do test the robustness of the offset to the addition of the MS template, discussed later in this section.

We use the same superpixel and subpixel resolution as in the analysis of the simulations. We apply the HMC sampler as discussed in \cref{sect:2.2}. We use the same values for the HMC sampler hyper-parameters (Leap-Frog step sizes and the number of jumps) as were used in the simulations for the analysis of the Planck data. We obtain $2 \times 10^4$ samples of the emissivity and the offset parameters over MaskHI6, discard the first $10^{3}$ samples, and perform the analysis with the remaining $1.9 \times 10^{4}$ samples. Using the remaining $1.9 \times 10^{4}$ MC samples, we compute the posterior mean and standard deviation of the model parameters. The Gelman-Rubin convergence diagnostics values for the Planck 353 GHz data estimated using seven independent chains are 1.0003 (for emissivity) and 1.003 (for offset). This shows that the chains are converged.

\begin{figure}
\centering
\includegraphics[width=\columnwidth]{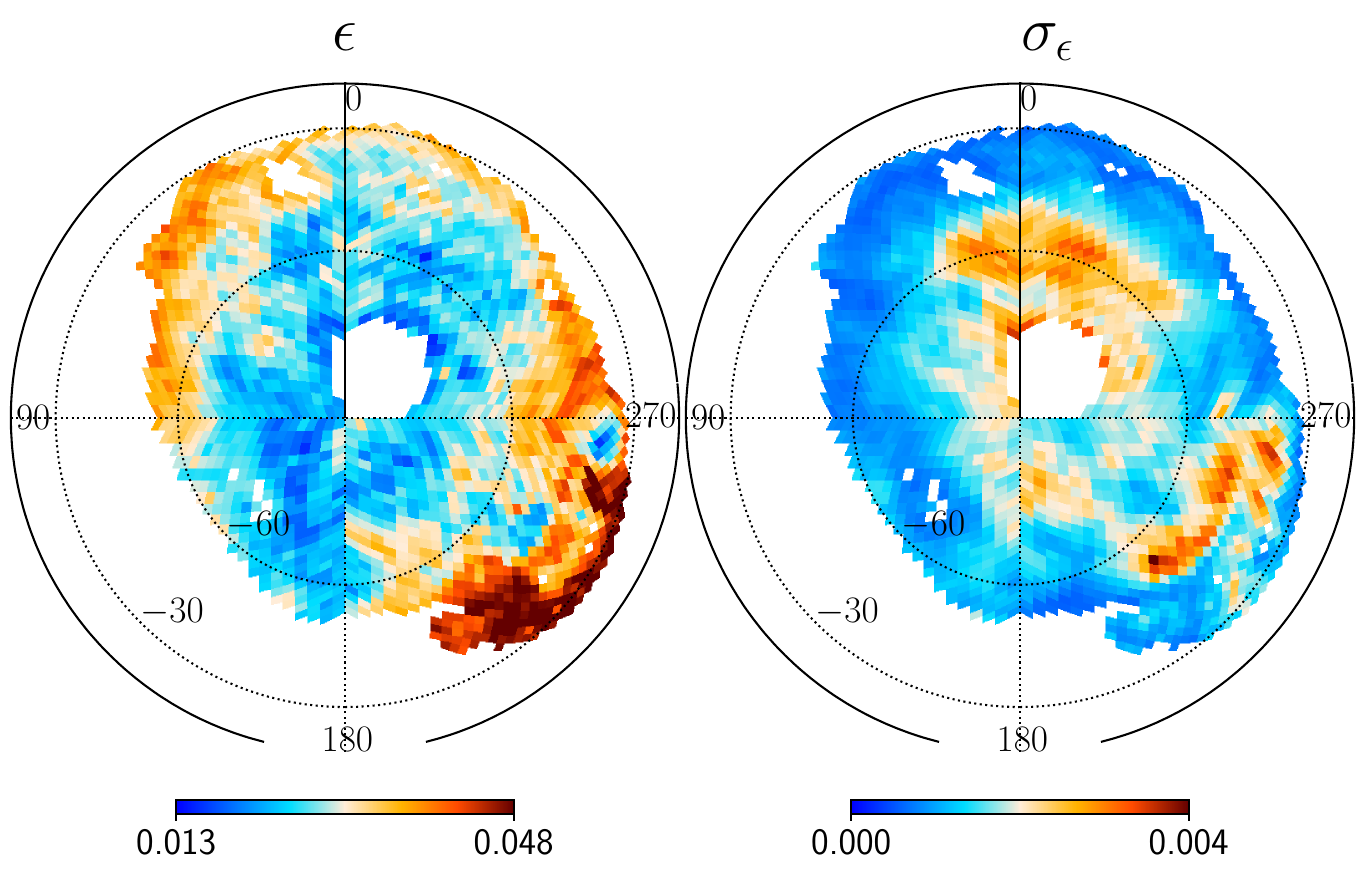}
\caption{The mean (\emph{left}) and the standard deviation (\emph{right}) of the dust emissivity obtained from the Planck 353 GHz intensity map. Both the maps are in $\MJysr (10^{20} \mathrm{cm^{-2}})^{-1}$ units. These are for the analysis with the MaskHI6 mask, which has $f_{\rm sky} = 15.3\%$.} 
\label{fig:emissivity_353GHz}
\end{figure}
\begin{figure}
\centering
\includegraphics[width=\columnwidth]{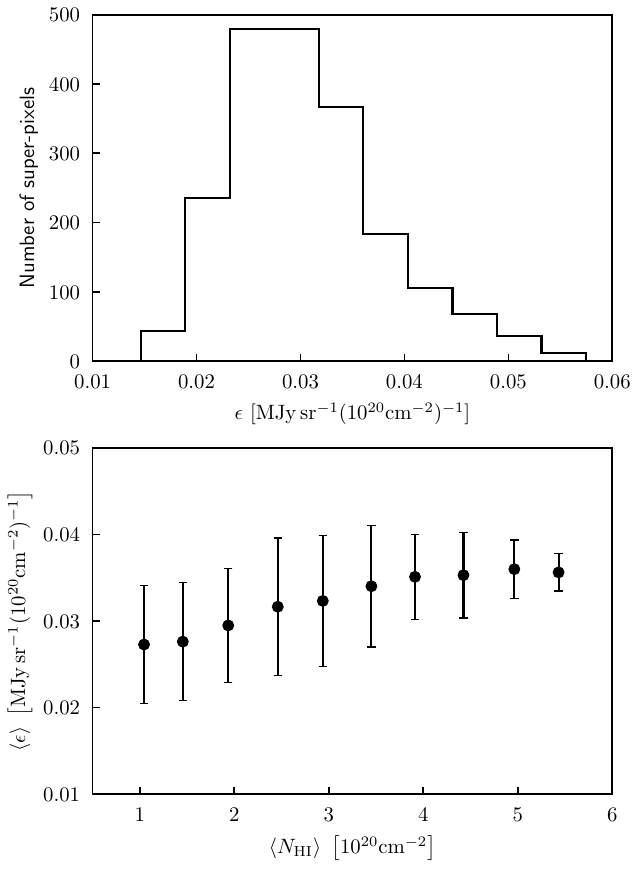}
\caption{\emph{Upper panel}: Histogram of mean dust emissivity obtained in the Planck 353 GHz intensity map analysis over MaskHI6. The map of these dust emissivity values is given in the \emph{left panel} of \cref{fig:emissivity_353GHz}. \emph{Lower panel}: Variation of dust emissivity as a function of mean dust column density. $\langle\NHI\rangle$ represents the bin average \hi\ column density and $\langle\epsilon\rangle$ are the average over the superpixels within the same \NHI\ bins. The error bars are the standard deviations in the respective bins.} 
\label{fig:histogram_emissivity_353GHz}
\end{figure}

In \cref{fig:emissivity_353GHz}, we show the map of the mean and the standard deviation of the dust emissivity for analysis with the MaskHI6 mask. The mean $\epsilon$ map shows the fluctuations that extend down to the scales of superpixels, indicating small-scale variations (1.8\deg\ pixel size) in dust emissivity.  The standard deviation map shows the inhomogeneity in the uncertainty of the dust emissivity. The \NHI\ map partly determines the inhomogeneity in $\sigma_{\epsilon}$ map as the instrument noise and the CIB are fairly statistically isotropic. This map also allows us to assess the spatial variation in the dust emissivity. The spatial average of the dust emissivity map in \cref{fig:emissivity_353GHz} is $0.031 \,\MJysr (10^{20} \mathrm{cm^{-2}})^{-1}$ and the standard deviation of the dust emissivity values over the unmasked sky area is $0.007\,\MJysr (10^{20} \mathrm{cm^{-2}})^{-1}$. Note that, as shown in \cref{fig:histogram_emissivity_353GHz}, the spatial distribution of the dust emissivity is slightly skewed towards higher values. The upper panel of \cref{fig:histogram_emissivity_353GHz} shows the histogram of dust emissivity values in the $\epsilon$ map depicted in \cref{fig:emissivity_353GHz}. The lower panel of \cref{fig:histogram_emissivity_353GHz} shows the variation of dust emissivity as a function of \NHI. The data point and the error bar represent the mean and standard deviation of the emissivity computed over superpixels that fall within the given \NHI\ bin. The average dust emissivity increases with increasing \hi\ column density. Our spatial average value of the dust emissivity is $\approx 20\%$ lower than the value quoted in \citet{planck-XVII:2014}. This difference could arise due to (1) different aperture sizes considered for the dust-\hi\ correlation analysis in \citet{planck-XVII:2014} and (2) change in the calibration of the Planck 353 GHz map from PR1 to PR3. For MaskHI6, we report the global offset value equal to $0.1284\pm0.0003\,\MJysr$.

\begin{figure*}
\centering
\includegraphics[width=\linewidth]{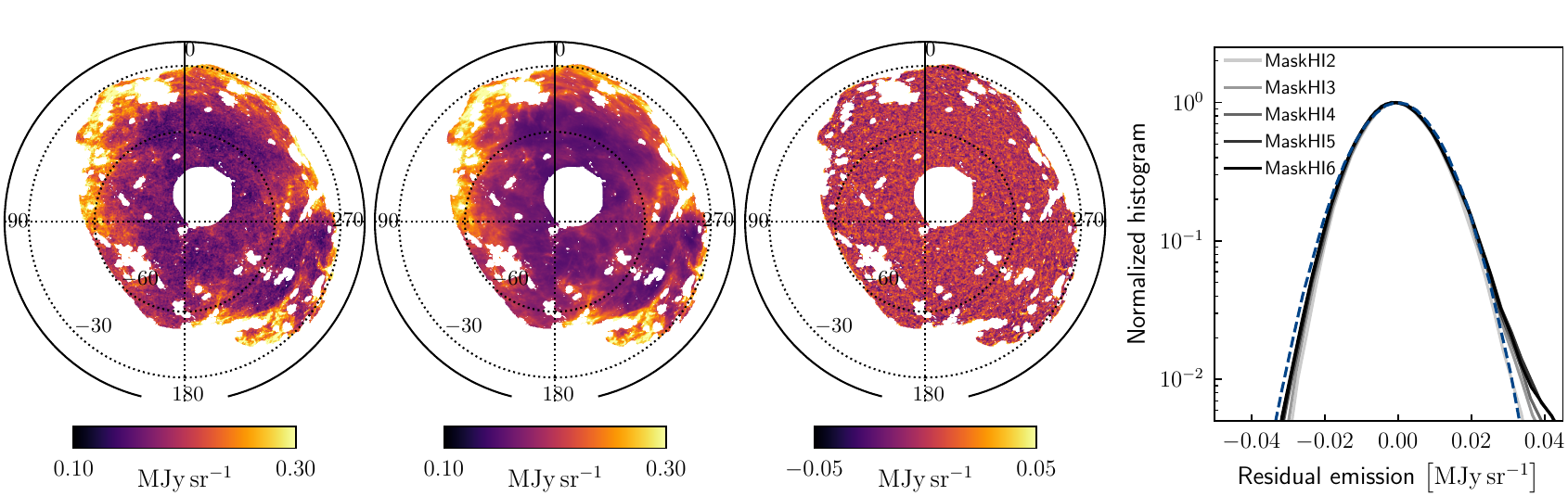}
\caption{Results of analysis with Planck data at 353 GHz in the \MJysr\ unit. \emph{First panel}: the CMB subtracted Planck intensity map at 353 GHz over the South Galactic Pole region, \emph{Second panel}: the estimate of the signal map obtained using the posterior mean emissivity and the offset, \emph{Third panel}: the residual map, which is the difference between the difference map and the signal estimate, \emph{Fourth panel}: histogram of the residual map for the analysis with different sky masks. The dashed blue curve shows the expected residual based on the CIB and the instrument noise.}
\label{fig:result_353GHz}
\end{figure*}

We obtain the signal model map using the mean dust emissivity map and the mean global offset. As superpixels do not overlap and the emissivity varies over the superpixels, the estimated mean dust emissivity map becomes discontinuous at the scale of \healpix\ superpixel with 1.8\deg\ angular size (pixel size at $\Nside = 32$). To avoid this discontinuity, we first obtain the dust emissivity map at $\Nside = 512$ by interpolating the $\Nside = 32$ pixel values using \texttt{interpolate.griddata} function of \texttt{scipy} and then smooth the resultant mean dust emissivity map with a Gaussian kernel of 1.8\deg\ FWHM. We construct the signal model map using the smoothed dust emissivity map and the mean value of the global offset following \cref{eq:model}. We show the results of this analysis in \cref{fig:result_353GHz}. The leftmost panel in \cref{fig:result_353GHz} shows the CMB-subtracted Planck intensity map at 353 GHz over the extended unmasked region eMaskHI6, and the second panel shows the signal map. The third panel in \cref{fig:result_353GHz} shows the difference between the CMB-subtracted Planck 353 GHz intensity map and signal model map over the same sky mask. 

To compare the residual distribution with the expected distribution of the residual at 353 GHz, we simulate a random Gaussian realization of the CIB at a beam resolution of 16.2\arcm\ (FWHM) using the best-fit CIB model angular power spectrum and an uncorrelated Gaussian white noise realization from the E2E smoothed noise variance map and add them. The expected distribution of the total noise is centred around zero because both the CIB and the instrumental noise have a zero mean. If our model assumptions are correct, the distribution of the residual should be consistent with the distribution of the expected fluctuations from the CIB and instrument noise, similar to as seen in the simulated data (see Figure~\ref{fig:8_sec4}). The fourth panel of \cref{fig:result_353GHz} shows the residual for masks with different sky fractions and compares it with the residual expected from the CIB and the instrument noise for the respective sky fraction. Though the residual distributions obtained using different MaskHIQ masks are mostly consistent with the expected model residuals from the CIB and instrumental noise contributions, they have a minor fraction of pixels contributing to the non-Gaussian tails with increasing sky fractions. The residual histogram is in logarithmic scale to highlight the asymmetric nature of the positive tail arising from a very small number of localised pixels in which dust is uncorrelated with the Galactic \hi.

\cref{tab:data-mask} shows the recovered offset and recovered mean emissivities at 353 GHz over different sky masks. As the available sky area decreases, the mean value of dust emissivity decreases, indicating a \hi\ column density dependence of dust emissivity. We also see an opposite trend in the global offset value, which increases with the reduction in the sky area and decreasing average emissivity. This is expected as a decrease in the global offset value compensates for an increase in the mean dust emissivity.
Over the range of \NHI\ thresholds and corresponding sky area, the inferred offset parameter ranges from 0.1358 \MJysr\ (over MaskHI2) to 0.1284 \MJysr\ (over MaskHI6). This range encompasses the value of the CIB monopole in the 353 GHz intensity map. The offset we infer is the total zero-level of the map that includes the CIB or any other residual zero-level. The differences between the CIB monopole value and the value we infer could be due to the differences in (1) the smoothing scale of the Planck and \hi\ column density map, (2) the Planck Galactic zero-level is estimated over a larger sky fraction (28\% of the sky) than considered here, and (3) the difference in the local velocity cloud \hi\ column density threshold ($ < 3 \times 10^{20}$\,cm$^{-2}$) from the LAB survey \citep{Kalberla:2005}. 


To check the robustness of the global offset value due to the modelling error, we perform the inference with the two-template fit using Galactic \hi\ and MS templates using the methodology discussed in \cref{subsubsec:multiple_templates}. We infer the offset $O = 0.1281\pm0.0003 \MJysr$ at 353 GHz over MaskHI6 for the two-template fit. The average of best-fit values of recovered emissivities for Galactic \hi\ and MS templates are  $0.031\,\MJysr (10^{20} \mathrm{cm^{-2}})^{-1}$ and $-0.6 \times 10^{-4} \MJysr(10^{20} \mathrm{cm^{-2}})^{-1}$, respectively. The corresponding $1\sigma$ deviation on them are $0.007\,\MJysr (10^{20} \mathrm{cm^{-2}})^{-1}$ and $1.8\times 10^{-4} \MJysr(10^{20} \mathrm{cm^{-2}})^{-1}$, respectively.

\begin{table*}
\caption{Results of analysis of Planck intensity map at 353 GHz. We list the global offset ($O$) values and the mean dust emissivity over the superpixels ($\langle\epsilon^{j}\rangle$) estimated over different sky masks corresponding to different \NHI\ cutoffs. Column $\sigma$ represents the uncertainty on the mean dust emissivity $\langle\epsilon^{j}\rangle$. We also include the estimates from the Planck collaboration analysis for comparison.}  
\label{tab:data-mask}
\begin{tabular} { |  c | c c c|}
\hline 
  \thead{Sky masks\\($f_{\rm {sky}}\left[\%\right]$)} &  \thead{$O$ \\ $\left[\mathrm{MJy\,sr^{-1}}\right]$} &  \thead{$\langle \epsilon^{j} \rangle$\\$\left[\mathrm{MJy\,sr^{-1}}(10^{20} \mathrm{cm^{-2}})^{-1}\right]$} & \thead{$\sigma$ \\ $\left[\mathrm{MJy\,sr^{-1}}(10^{20} \mathrm{cm^{-2}})^{-1}\right]$}\\
  \hline
  MaskHI2 (7.3) & $0.1358 \pm 0.0005$ & $0.023$ & $0.00022$\\
  MaskHI3 (11.5) & $0.1332 \pm 0.0004$ & $0.026$ & $0.00019$\\
  MaskHI4 (13.9)& $0.1311 \pm 0.0004$ & $0.029$ & $0.00018$\\
  MaskHI5 (15.0)& $0.1294 \pm 0.0003$ & $0.030$ & $0.00017$\\
  MaskHI6 (15.3)& $0.1284 \pm 0.0003$ & $0.031$ & $0.00016$\\
\hline
\thead[c]{Planck analysis\\\citep{planck-XVII:2014}\\\citep{planck-I:2018}} & $0.130\pm 0.038$ & $0.039$ & $0.0014$   \\
\hline
\end{tabular}
\end{table*}

\section{Summary}\label{sec:summary}

We demonstrate the use of the Bayesian inference method to estimate the pixel-dependent dust emissivity and the global offset in the diffuse interstellar medium at far-infrared and submillimeter frequencies. We utilise the dust-\hi\ correlation to model the Galactic thermal dust emission, using the integrated Galactic \hi\ column density map from the GASS survey \citep{GASSI:2009} as a template.  The signal model incorporates spatially varying dust emissivity and global offset over a 6300 deg${}^2$ region centred around the southern Galactic pole. The HMC method allows efficient sampling of the high dimensional posterior distribution of the dust emissivity and the offset parameters.

We first validate the method on the  Planck simulations, which include the CIB and instrumental noise. The dust emissivity parameters are fixed based on earlier Planck analysis \citep{planck-XVII:2014}. This validation process allows us to test the analysis pipeline and fix the parameters of the HMC sampler that we use for the real Planck data. Given that the data is on the partial sky, the inference of the offset can be biased if any residual CMB dipole is not taken into account. We demonstrate the nature of the inferred parameters in the presence of residual CMB dipole at 217 GHz. We show that the amplitude of the residual CMB dipole is highly correlated with the global offset parameter for partial sky analysis. A small residual CMB dipole does not significantly affect the global offset inference beyond the increase in the error bar of the global offset, whose uncertainty is still significantly smaller than the global offset value at 217 GHz. At 353 GHz, the error bar on the fitted residual CMB dipole term increases due to increased noise contributions from the CIB and the instrumental noise. However, we note that the joint dipole and offset inference will be important in the case of application to the NPIPE data where the CMB dipole is retained in the frequency maps \citep{Planck:2020olo}. We further test the robustness of the inferred offset in the presence of multiple \hi\ templates in the signal model. We consider the emission from the MS as an additional component in the signal model. The global offset value does not change significantly when considering the two-template analysis --- Galactic \hi\ and MS templates. 

As a demonstration of the method, we apply the same HMC methodology to the Planck intensity map at 353 GHz. Results of this analysis are depicted in \cref{fig:emissivity_353GHz}, \cref{fig:histogram_emissivity_353GHz}, \cref{fig:result_353GHz}, and \cref{tab:data-mask}. As shown in \cref{fig:emissivity_353GHz}, we infer the emissivity of the dust over the sky area $\approx 7500$ deg$^2$ and its variation at scales larger than 1.8\deg. For the region of interest over $7500$ deg$^2$ area centered around the southern Galactic pole with \NHI\ threshold $\NHI < 6 \times10^{20} \text{cm}^{-2}$, the spatial average of dust emissivity is $0.031~\mathrm{MJy\,sr^{-1}}(10^{20} \mathrm{cm^{-2}})^{-1}$ with standard deviation of $0.007~\mathrm{MJy\,sr^{-1}}(10^{20} \mathrm{cm^{-2}})^{-1}$. The inferred offset for MaskHI6 is $0.1284\pm 0.0003\,\MJysr$ is close to the monopole of the \citet{B_thermin:2012} CIB model value $0.130 \pm 0.038\,\MJysr$ added to the Planck intensity map at 353 GHz after setting the Galactic zero level \citep{planck-I:2018}. We further performed the same analysis on the smaller sky mask by putting lower \NHI\ thresholds. As expected, we find the inferred value of the global offset is stable for different sky masks. We also find that the mean value of the dust emissivity decreases as we go to the low column density regions or lower sky masks. The non-Gaussian tail in the residual distribution (caused by the emission not correlated with \NHI) does depend on \NHI\ threshold and sky fraction. This additional emission component can be partly dealt with as an additional noise component in the data model, denoted by $I^{\rm R}_{\nu}$ in \cref{eq.1}, which we leave for future work.

The methodology introduced in this paper opens a new way to jointly estimate the pixel-dependent dust emissivity and a global offset over the field of interest using the dust-\hi\ correlation. The HMC method is able to sample around $2\times 10^3$ parameters and infer the parameter posterior distribution. This method can be useful in studying the 3D variation of dust SEDs to constrain the frequency decorrelation of dust B-modes. In the future study, we will apply this technique to study the dust emissivity properties associated with different \hi\ phases (CNM, LNM, and WNM) of diffuse ISM considered in \cite{Adak:2019} and \cite{T_Ghosh:2017}, which will be useful to extend the sky model of dust polarisation at multiple sub-mm frequencies. The residual maps estimated using this method will be useful for estimating the CIB maps at the best angular resolution of Planck. In the forthcoming paper, we will apply the same methodology to other Planck HFI frequency maps to separate the thermal dust emission from the CIB emission and characterise the dust emissivity parameters.


\section*{Acknowledgements}
The Planck Legacy Archive (PLA) contains all public products originating from the Planck mission, and we take the opportunity to thank ESA/Planck and the Planck Collaboration for the same. 
This work has used \hi\ data of the GASS survey headed by the Parkes Radio Telescope, a part of the Australia Telescope funded by the Commonwealth of Australia for operation as a National Facility managed by CSIRO. Some of the results in this paper have been derived using the \healpix\ package. All the computations in this paper were run on the Aquila cluster at NISER. This work was supported by the Science and Engineering Research Board, Department of Science and Technology, Govt. of India, grant number SERB/ECR/2018/000826. D. \ Adak acknowledges UGC for providing Ph.D. fellowship when a major part of this project is done. D. \ Adak acknowledges IMSc for providing a postdoc fellowship for one year when the rest of the project is completed. S.\ Shaikh acknowledges support from SERB for the Research Associate position at NISER, during which a part of this work was performed, and the Beus Center for Cosmic Foundations for current support. S.\ Sinha would like to thank NISER Bhubaneswar for the postdoctoral fellowship.  G. \ Lagache acknowledges funding from the European Research Council (ERC) under the European Union’s Horizon 2020 research and innovation programme (grant agreement No 788212) and from the Excellence Initiative of Aix-Marseille University-A*Midex, a French “Investissements d’Avenir” programme.

\section*{DATA AVAILABILITY}
The Planck 353 GHz intensity map and the component-separated \smica\ CMB map are available at \url{https://pla.esac.esa.int}. The Galactic \hi\ column density map is downloaded from \url{https://www.astro.uni-bonn.de/hisurvey/index.php}. The different sky masks, the dust model map and the residual map at 353 GHz generated in this analysis are publicly available at \url{https://github.com/shabbir137/dust_emissivity.git}. 




\bibliographystyle{mnras}
\bibliography{sed} 


\appendix

\section{Details of HMC implementation} \label{app:HMC_details}
Different MCMC sampling methods differ from each other mainly in the way they generate the proposed point. 
HMC uses Hamiltonian dynamics to generate the proposed point \citep{Duane:1987de, Neal:2012}. This is accomplished by introducing an additional set of parameters, called \textit{momentum parameters}, one momentum parameter ($p_i$) corresponding to each parameter of interest ($q_i$). The momentum variables are chosen to follow a Gaussian distribution with a covariance defined by a \textit{mass matrix}. One has to choose the mass matrix specific to the problem. This aspect is similar to the choice of parameters of the proposal distribution in sampling with the Metropolis-Hastings algorithm. The multidimensional Gaussian distribution of momenta is augmented with the original probability distribution that needs to be sampled. For example, if we have `$q_i$' as $n$ number of parameters of interest with the probability distribution $\calP(\{q_i\})$, the probability distribution that is sampled in HMC is the joint probability distribution of ${p_i}$ and ${q_i}$:
\begin{eqnarray}
 \calP(\{q_i, p_i\}) &\equiv& \frac{\exp(-\calH)}{(2 \pi)^{D/2} \sqrt{|\boldsymbol{\mu}|} } \\
 &=& \frac{1}{(2 \pi)^{D/2}\sqrt{\boldsymbol{|\mu}| } } \exp\Big [-\frac{\textbf{p}^T \boldsymbol{\mu}^{-1} \textbf{p}}{2}  \Big] \calP(\{q_i\}). \nonumber
\end{eqnarray}
Here, $\textbf{p}$ is the vector of momenta $\{p_i\}$ associated with the parameters and $\calH$ is the \textit{Hamiltonian} comprising of a \textit{Kinetic Energy} term and a \textit{Potential Energy} term: 
\begin{equation}
\calH(\{q_i, p_i\}) = \frac{1}{2} \textbf{p}^T \boldsymbol{\mu}^{-1} \textbf{p} - \ln[\calP(\{q_i\})] \ .
\end{equation}
In the above equation, the first term on the right-hand side is the kinetic energy term, and the second term, $- \ln[\calP(\{q_i\})]$, acts as potential energy. $\boldsymbol{\mu}$ is the mass matrix corresponding to the set of parameter $\{ q_i \}$. In general, $\boldsymbol{\mu}$ is chosen as equal to the inverse of the covariance matrix of the parameters of interest. This choice may lead to $\boldsymbol{\mu}$ having off-diagonal elements, leading to more computational cost. Instead, we choose the diagonal mass matrix. Further details about this choice and its implications are discussed later in this section. With a diagonal mass matrix, the resulting Hamiltonian is
\begin{equation}
\calH(\{q_i, p_i\}) = \sum_{i} \frac{p_i^2}{2\mu_i} - \ln[\calP(\{q_i\})] \ .
\end{equation}

Hamilton's equations for variable $q$ and the conjugate momentum $p$ are
\begin{equation}
 \dot{q} \equiv \frac{dq}{dt} = \frac{\pd \calH}{\pd p} \quad \text{and} \quad
 \dot{p} \equiv \frac{dp}{dt} = -\frac{\pd \calH}{\pd q}.
\end{equation}
The above equations are solved using symplectic integration methods. In this work, we choose the Leap-Frog method to simulate the Hamiltonian dynamics. 
\begin{equation}
    p(t + \Delta/2) = p(t) + \frac{\Delta}{2} \dot{p}(q(t));
\end{equation}
\begin{equation}
    q(t + \Delta) = q(t) + p(t + \Delta/2) \Delta/\mu;
\end{equation}
\begin{equation}
    p(t + \Delta) = p(t + \Delta/2) + \frac{\Delta}{2} \dot{p}(q(t+\Delta)).
\end{equation}
In the above equations, $\dot{p}(q(t))$ is the momentum derivative substituted from \cref{P_dot_eps_rev} or \cref{P_dot_Onu_rev} depending on the parameter.

Below is the schematic of the algorithm we followed in sampling the posterior distribution. The algorithm is written in terms of the generic variables $q_i$ (which refer to $\{ \epsilon^{j,t}_{\nu}, \, O_{\nu},\, a^{(R/I), \nu}_{1m}\}$) and $p_i$.
\begin{algorithmic}[1]
\State Initialise $q^{(0)}$
\For{$k=1...N_{s}$ $\vphantom{\big[^\dagger}$ }
\State $p \sim \calN(0, \mu)$; where $\calN$ is Gaussian distribution with variance $\mu$.
\State $q_{(0)}^*, p_{(0)}^* = q^{(k-1)}, p$
\For{$j=1...N$}
\State a Leap-Frog move from $(q^{*}_{(j-1)}, p^{*}_{(j-1)})$ to $(q^{*}_{(j)}, p^{*}_{(j)})$ 
\EndFor 
\State $q^{*}, p^{*} = q^{*}_{(N)}, p^{*}_{(N)}$
\State Accept $q^{*}, p^{*}$ with the probability $\rm{min} (1, e^{-H(q^*, p^*) + 
H(q^{(k-1)}, p^{(k-1)})})$. If the proposed point is rejected, then $q^{k}, p^{k} = q^{k-1}, p^{k-1}$  
\EndFor $\vphantom{\big[^\dagger}$
\end{algorithmic}

As is common in the MCMC methods, we start the algorithm with a reasonable guess of the parameter values $\{q^{(0)}_i\}$. The corresponding momenta $\{p^k_i\}$ are drawn from a multi-normal distribution at the start of each $k^{th}$ step, including the very first step. With this as the initial point on the phase space trajectory, Hamilton's equations evolve these variables in the phase space. This evolution is simulated by $N$ Leap-Frog jumps to arrive at a proposed point in the phase space $q^*, p^*$. The leap-frog method is volume-preserving and time-reversible. However, Hamiltonian $(\calH)$ may not be conserved in practical numerical computations due to leapfrog discretization, which can introduce a bias with respect to the target distribution. This bias can be reduced with a sufficiently small step size, which leads to less discretization error. To avoid this bias, we use the Metropolis rule, according to which the proposed point is accepted with the probability equal to the minimum of $(1, e^{-H(q^*, p^*) + H(q^{(k-1)}, p^{(k-1)})})$.

Once the samples of $(p_i, q_i)$ are obtained, getting the Monte Carlo samples of $q_i$, which samples the distribution $\calP(q_i)$, is straightforward. Discarding the samples of $p_i$ from the joint samples of $(p_i, q_i)$ leads to the marginalisation of the distribution $\calP(q_i, p_i)$ with respect to $p_i$. As $p_i$ and $q_i$ are independent variables, the resultant samples of $q_i$ are the desired samples of $q_i$ drawn from the probability distribution of our interest $\calP(q_i)$.

\subsection{Burn-in, Correlation Length, and Convergence Test}\label{sec:5.2}
Here, we discuss some considerations for analysing these Markov chains before using these chains to draw the inference.
\paragraph*{Burn-in:}
We neglect some of the initial samples from the chain as Burn-in. To quantitatively determine the Burn-in sample size, we monitor the $\chi^2$ using a model map estimated with the increasing number of samples in the chains of the parameters. The model map was built from the posterior mean of the parameter chain with an increasing number of samples. Burn-in sample then consists of the initial points where $\chi^2$ is away from the total number of degrees of freedom. In practice, we discard the first $10^{3}$ samples, which is much larger than the Burn-in sample determined using $\chi^2$ criteria. Note that, in the absence of explicit priors on the parameters and Gaussian likelihood, $\chi^2$ is equal to the logarithm of the posterior up to a constant term.
\paragraph*{Correlation length:}
 When we draw the samples from the target distribution, all the samples may not be independent because of the correlation within a chain of samples. The effective number of the independent samples out of the total samples, $N_s$, is $n_{\rm eff} = N_s/L$, where \textit{correlation length}, $L$, is defined as,
\begin{equation}\label{eq.5.2.1}
    L = 1 + 2\sum_{t = 1}^{t_{\rm max}} \rho(t),
\end{equation}
where $\rho(t)$ is the auto-correlation coefficient of the chain for lag $t$ \citep{Taylor_Cl_with_HMC_2008}. 

Correlation length depends on the total jump, $s = N \Delta$ in the Leap-Frog scheme and the scale of the distribution of the parameters of the problem. The total jump $s$ should be of the order of scale of distribution of the parameters to get less correlated samples. $\Delta$ controls the accuracy with which we implement Hamiltonian dynamics. This, in turn, determines the acceptance rate of the proposed sample. For a given $\Delta$, $N$ determines how far the proposed sample is from the current position. If $\Delta$ is chosen to be too small, $N$ needs to be large enough to move a sufficient distance along the trajectory, resulting in increased computation time. In contrast, for the choice of a large value of $\Delta$, the computation of the phase space trajectories becomes relatively less accurate, resulting in a reduced acceptance rate. If $N$ is chosen to be small, samples are correlated, whereas too large a value of $N$ may bring back the proposed sample very close to the starting point after completing Leap-Frog jumps \citep{Neal:2012}. In \cref{app:var-nsteps}, we show the difference in correlation length for fixed and variable Leap-Frog jumps. The accuracy of Hamiltonian dynamics can be increased by efficient choice of total jump in the Leap-Frog scheme \citep{2011arXiv1111.4246H, nawaf2017aap, hoffman2021proced, sountsov2022ar}.

As discussed in \cref{sect:2.2}, we choose the mass matrix $\boldsymbol{\mu}$ to make step size $\Delta$ independent of the scale of the target distribution. In the approximations considered in our analysis, neglecting the correlation between subpixels belonging to two different superpixels renders the mass matrix sparse and diagonal dominant. The only non-zero off-diagonal terms are those due to the cross-correlation among emissivity, offset and spherical harmonic coefficients (see \cref{sect:2.2}). The rest of the off-diagonal terms are the cross-correlation between emissivities at different superpixels, which are zero due to the approximations considered in our analysis. Inverting the mass matrix and evolving the vector forms of the Leap-Frog equations with $\boldsymbol{\mu}^{-1}$ are relatively computationally costly when dealing with the non-diagonal mass matrix. Therefore, we neglect the off-diagonal terms in the column (and row) of the mass matrix that connect the emissivity and offset parameters. As a consequence of this choice, we have to choose two separate step sizes, $\Delta_{\epsilon}$ and $\Delta_{O}$, for emissivity and offset, respectively. For the various schemes used to tune the HMC hyper-parameters, see \citep{2011arXiv1111.4246H, nawaf2017aap, hoffman2021proced, sountsov2022ar} and for their implementation, refer to the probabilistic programming frameworks such as STAN \citep{JSSv076i01}, PyMC \citep{2015arXiv150708050S}, and pyro \citep{bingham2019pyro}.
\paragraph*{Convergence test:}
To quantify the convergence of the Monte Carlo sample, we adopt the Gelman-Rubin test \citep{Gelman-Rubin:1992}. To implement this test, one needs multiple Monte Carlo chains starting with sufficiently separated positions in the parameter space. One then computes the following ratio $R$, which should be ideally equal to one.
\begin{equation}\label{eq.5.2.2}
R = \frac{V}{W},
\end{equation}
where $V$ is the variance of {the given parameter} between the chain and $W$ is the variance of the same parameter along the chain \citep{Brooks:1998,Alan:2009}. In practice, it is recommended that $R$ should be less than 1.01 to consider the sample as converged to the distribution being sampled \citep{Vehtari:2019}.

\section{Variable Leap-Frog steps} \label{app:var-nsteps}
\begin{figure}
\centering
\includegraphics[width=\columnwidth]{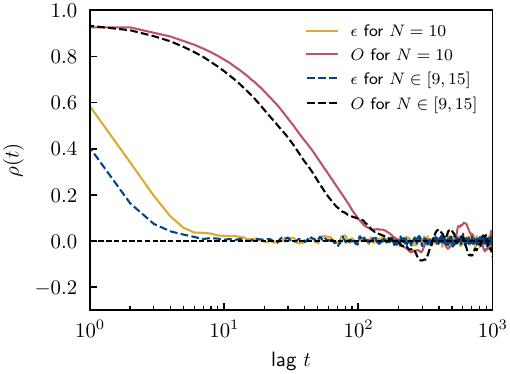}
\caption{Same as \cref{fig:4_sec4}, but for fixed and variable values of Leap-Frog jumps $N$. For a fixed $N = 10$, the correlation length for $\epsilon$ and $O$ are $10$ and $102$, respectively. For variable $N \in \left[9, 15\right]$, the correlation length for $\epsilon$ and $O$ are $7$ and $57$, respectively.}
\label{fig:corr-length-var}
\end{figure}
To investigate the change in correlation length with Leap-Frog jumps $N$, we perform the HMC sampling with variable $N$. Following \cite{Neal:2012}, we randomly draw $N$ at each iteration from a uniform distribution, $\mathcal{U}\left[9,15\right]$ and apply the HMC sampler on the simulated map at 353 GHz over MaskHI6 as discussed in \cref{sect:2.2}. We then compute the compute the auto-correlation coefficient of emissivity $\epsilon$ at one superpixel and the global offset $O$ as in \cref{sec:5.3}. \cref{fig:corr-length-var} shows the auto-correlation coefficient for fixed Leap-Frog jumps $N =10$ (\cref{sec:5.3}) and variable Leap-Frog jumps $N \in \left[9,15\right]$. It shows that for variable $N$, the correlation length decreases considerably for $O$. Variable $N$ minimises the correlation among the various global parameters for multiple template fit or fit with dipole contribution.

\label{lastpage}
\end{document}